\documentclass[usenatbib]{mnras}

\usepackage[T1]{fontenc}
\usepackage{ae,aecompl}

\usepackage{graphicx}	
\usepackage{amsmath}	
\usepackage{amssymb}	
\usepackage{bm}
\usepackage{caption}
\usepackage{url}
\newcommand{\Hii}{H\textsc{ii} }
\defcitealias{Metha+21}{M21}
\newcommand{\GeoGalsI}{\citetalias{Metha+21}}
\newcommand{\gradZ}{ \left\langle \nabla Z \right\rangle}
\newcommand{\NSH}{N$_2$S$_2$H$\alpha$}
\newcommand{\NO}{N$_2$O$_2$}
\newcommand{\ON}{O$_3$N$_2$}
\newcommand{\RS}{RS$_{32}$}

\title[Geostatistics of galaxies II]{A geostatistical analysis of multiscale metallicity variations in galaxies [II]: Predicting the metallicities of \Hii and diffuse ionised gas regions via universal kriging}

\author[Metha et al.]{
Benjamin Metha$^{1,2}$\thanks{\hbox{methab@student.unimelb.edu.au}}, Michele Trenti$^{1,2}$, Tingjin Chu$^{3}$, Andrew Battisti$^{2,4}$
\\
$^1$School of Physics, The University of Melbourne, VIC 3010, Australia\\
$^2$Australian Research Council Centre of Excellence for All-Sky Astrophysics in 3-Dimensions, Australia\\
$^3$School of Mathematics and Statistics, The University of Melbourne, VIC 3010, Australia\\
$^4$Research School of Astronomy and Astrophysics, Australian National University, Cotter Road, Weston Creek, ACT 2611, Australia
}

\date{Accepted XXX. Received YYY; in original form ZZZ}

\pubyear{2021}

\begin{document}
\label{firstpage}
\pagerange{\pageref{firstpage}--\pageref{lastpage}}
\maketitle

\begin{abstract}
The metallicity of diffuse ionised gas (DIG) cannot be determined using strong emission line diagnostics, which are calibrated to calculate the metallicity of \Hii regions. Because of this, resolved metallicity maps from integral field spectroscopy (IFS) data remain largely incomplete.
In this paper (the second of a series), we introduce the geostatistical technique of universal kriging, which allows the complete 2D metallicity distribution of a galaxy to be reconstructed from metallicities measured at \Hii regions, accounting for spatial correlations between nearby data points.
We apply this method to construct high-fidelity metallicity maps of the local spiral galaxy NGC 5236 using data from the TYPHOON/PrISM survey. We find significant correlation in the metallicity of \Hii regions separated by up to $0.4-1.2$ kpc. 
Predictions  constructed using this method were tested using cross-validation in \Hii regions, and we show that they outperform significantly interpolation based on metallicity gradients. 
Furthermore, we apply kriging to predict the metallicities in regions dominated by DIG emission, considering seven additional spiral galaxies with high resolution ($\lesssim 100$pc) metallicity maps. We compare kriging maps to DIG metallicities computed with novel ionisation corrections, and find that such corrections introduce a systematic offset of up to $\pm0.1$ dex for any individual galaxy, with a scatter of $0.02-0.07$ dex for the sample.
Overall we recommend universal kriging, together with a calibrated geostatistical model, as the superior method for inferring the metallicities of DIG-dominated regions in local spiral galaxies, demonstrating further the potential of applying geostatistical methods to spatially resolved galaxy observations.

\end{abstract}



\section{Introduction} \label{sec:intro}
  
 Recent advances in integral field spectroscopy (IFS) have produced datacubes containing an unprecedented amount of information about nearby galaxies, revealing how stellar masses, ages, metallicities, and luminosities, gas metallicities and kinematics, and dust extinction vary spatially. These maps offer increasingly high resolution, but are also potentially affected by missing data and/or regions where the signal to noise ratio is degraded. In such big and complex data frameworks, traditional astronomical analysis techniques are not necessarily suited to fully extract the wealth of information that these data products contain.

One particularly information-rich data-product that can be extracted from IFS surveys is the distribution of metals (elements heavier than helium) throughout the interstellar medium (ISM) of a galaxy. How the metallicity varies spatially throughout a galaxy is a product of many competing factors,
including (i) continuous metal enrichment throughout the star-formation history of a galaxy \citep[e.g.][]{EdmundsGreenhow95, Freeman+Bland-Haworth02}, (ii) turbulent mixing driven by thermal or gravitational instabilities \citep{deAvillez+MacLow02, Scalo+Emlegreen04, Pan+Scan10, KT18} or large structures such as spiral arms or bars \citep{DiMatteo+13, Grand+16, Ho+17}, (iii) galactic outflows caused by supernovae or galactic winds \citep{Heckman+90, Hopkins+14, Christensen+18}, and (iv) inflows of pristine gas \citep{Sanchez-Almeida+14} or previously ejected materials (galactic fountains, \citealt{Kim+20}). Understanding the relative strengths, spatial scales, and timescales of these processes is a major challenge in modelling galaxy formation and evolution \citep{Naab+17}.

Geostatistics is a subfield of spatial statistics that focuses on understanding stochastic processes that occur over a continuous spatial domain. As both star formation and the turbulent processes that mix the products of star formation into the surrounding ISM are stochastic processes, and galaxies in the local Universe are viewed as continuous extended objects, this branch of mathematics can be naturally applied to the field of extragalactic astronomy, offering a novel perspective on galaxy evolution.

Already, these techniques have been successfully applied to improve traditional data analysis recipes and extract interesting information from IFS data. \citet{Gonzalez-Gaitan+19} showed that geostatistical methods hold the potential to offer improvements over binning techniques (such as the popular Voronoi tessellation, used by e.g. \citealt{Cappellari+Copin03}) for accurately predicting the age and metallicity of stellar populations in IFS data without sacrificing spatial resolution. In order to capture non-symmetric deviations around a mean metallicity gradient, \citet{Clark+19} fit a geostatistical model using Gaussian process regressions to accurately map the dust mass-to-light ratio throughout two nearby galaxies. Widely used in machine learning algorithms \citep{Rasmussen+Williams06}, these Gaussian process regressions attempt to reconstruct a random field by creating a model of the covariance between nearby data points, allowing for complex structures to be captured with only a few parameters. These techniques were further employed by \citet{Williams+22}, who found that 12 out of 19 galaxies observed as a part of the PHANGS-MUSE survey showed significant 2D metallicity variations that could not be captured by a simple metallicity gradient model.
\citet{Metha+21} (hereafter \GeoGalsI) used a semivariogram\footnote{A \textit{semivariogram} is a geostatistical function that shows how the variance between data points in a sample depends on their spatial separation.} analysis to distinguish these small-scale metallicity fluctuations from measurement error, allowing the predictions of an analytical metal-mixing model to be directly compared to IFS data.

\textit{Kriging} refers to a family of statistically optimal, unbiased algorithms that can be used to fit a model of a Gaussian process to data while minimising prediction variance \citep{Cressie90}. Since their inception in the late 1960s \citep{UK}, these methods have formed a cornerstone of the geostatistical paradigm, seeing applications in a variety of fields, from environmental science and meteorology to economics and population statistics \citep{Wikle+19}. 

In this study (the second in a series), we continue the exploration on how geostatistical techniques may be used to improve the analysis of high-resolution IFS data, which started in \GeoGalsI. First, we construct a geostatistical hierarchical model of the multiscale metallicity structure of the warm ISM for NGC 5236, using mock-IFS data from the TYPHOON/PrISM survey. Then, using the method of universal kriging, we derive robust predictions, with uncertainty, about the metallicity in regions dominated by diffuse ionised gas (DIG), where the metallicity cannot be measured directly. Finally, we compare these DIG metallicity predictions to metallicities determined using the empirical correction factors of \citet{Kumari+19}, in order to offer an independent test of these new diagnostics.

This paper is organised as follows. In Section \ref{sec:DIG_review}, we review the main challenges in estimating DIG metallicity, before describing the TYPHOON survey in Section \ref{sec:data}. In Section \ref{sec:maths}, we introduce hierarchical geostatistical models and our model fitting methods before presenting the algorithm used for universal kriging. In Section \ref{ssec:M83_model}, we discuss in detail the results of our geostatistical model fitting procedure for the local spiral galaxy NGC 5236 (also known as M83), and confirm using 10-fold cross validation that our universal kriging method gives robust predictions for the metallicity of \Hii regions in Section \ref{ssec:validation}. After validating our methodology, we extend this analysis to DIG-dominated regions in Section \ref{sec:vs-dig}, providing an independent test of the DIG-corrected metallicity diagnostics of \citet{Kumari+19}. Strengths, weaknesses, and future applications of the universal kriging approach are discussed in Section \ref{sec:discussion}, and our main results are summarised in Section \ref{sec:conclusions}.

\section{Diffuse ionised gas metallicity determination} \label{sec:DIG_review}

There are several methods to measure metallicities within a galaxy's ISM  (see e.g. \citealt{Kewley+19, MaiolinoMannucci19} for comprehensive reviews). The abundance of an element can be measured directly from emission-line spectroscopy, if the temperature and density of the gas it resides inside is known \citep{Aller84}. However, the auroral lines required for accurate temperature measurements are often too faint to be seen in many systems of interest, as well as in IFS data \citep[e.g.][]{Andrews+Martini13}.
In these cases, strong emission-line calibrations from photoionisation models \citep[e.g.][]{Kewley+Dopita02, Dopita+16}, or empirical diagnostics \citep[e.g.][]{Pilyugin+Thaun05, Pilyugin+Grebel16} may also be used.

One limitation of all of these methods is that they are only valid for determining metallicities within \Hii regions, where the ISM has been completely ionised by local young, bright O/B stars. Narrow-band imaging of local galaxies targeting the H$\alpha$ line shows that approximately half of the H$\alpha$ flux emitted from the ISM originates from a thick disk of low density plasma that extends beyond the galactic mid-plane known as diffuse ionised gas (DIG) \citep{Zurita+2000, Oey+07}. Emission line spectra of the DIG are complex, revealing that it is $\sim 2000$K hotter than \Hii regions, ionised by a harder spectrum \citep{Hoopes+Walterbros03, Haffner+09}, and contains enhanced forbidden line ratios for [O \textsc{ii}]/H$\alpha$ and [N \textsc{ii}]/H$\alpha$. Many different sources for the ionisation of the DIG have been proposed, including radiation escaping from \Hii regions in the discs of galaxies \citep{Zurita+2000, Zurita+02}, radiation from evolved, low-mass stars \citep{Stasinska+08}, supernova-driven shocks \citep{Dopita+Sutherland95}, excitation within mixing layers between regions of hot and cool gas \citep{Binette+09}, or any combination of these phenomena \citep{Sanchez20, Mannucci+21}.

Despite the limited understanding of the ionisation sources within the DIG, it may still be possible to determine the metallicity within this gas using empirical means. Using data from the MUSE Atlas of Disks (MAD) survey \citep{Erroz-Ferrer+19}, \citet{Kumari+19} measured the difference in strong emission line ratios between \Hii regions and nearby DIG-dominated spaxels. Under the assumption that the ISM is homogeneous on spatial scales smaller than $0.5$ kpc, 
\citet{Kumari+19} searched for systematic offsets in the emission line ratios from pairs of \Hii and DIG-dominated spaxels 
and found that these offsets were correlated with the \mbox{[O \textsc{iii}]/H$\beta$} line ratio. 
Using this observation,
two corrections to popular emission-line diagnostics of \citet{Curti+17} were constructed
which may be used to more accurately infer the metallicity of DIG-dominated regions.

While the results of this approach are promising, there are two potential caveats that should be taken into account. Firstly, a recent geostatistical analysis revealed that the ISM of local spiral galaxies contain significant metallicity fluctuations on scales smaller than $\sim 1$kpc (\GeoGalsI), inconsistent with the assumption that the ISM is homogeneous on spatial scales smaller than $0.5$ kpc (although see \citealt{Kreckel+20} and \citealt{Williams+22} for different conclusions on the characteristic mixing scale).
Secondly, each of the aforementioned sources that ionise the DIG have their own spectral properties, and the relative contribution of these sources is not expected to be the same in different galaxies.
For this reason, it is unlikely that a single emission-line based correction factor exists that will be able to correct the DIG contamination stemming from all of these different components in all galaxies.

To overcome these issues we present here an alternate opportunity. The approach to estimating the metallicity of DIG-dominated regions presented in this study makes no assumptions about the ionisation sources of the DIG, depending instead on the spatial locations of the DIG/\Hii spaxels, and a model of metal mixing throughout galaxies. Under the assumption that the processes that govern turbulence and gas-mixing are constant throughout all parts of the ISM that are in the same phase, the predictions made by a geostatistical model trained on \Hii region data will be able to produce accurate predictions about the metallicity of all components of the warm interstellar medium.

\section{Data} \label{sec:data}

\begin{table*}
    \centering
\begin{tabular}{llllrrrr}
\hline
Name &  Type & RA & Dec & PA & i & D (Mpc) & $R_e$ (arcmin) \\
\hline
NGC 1068	&	Sb	&	02:42:40.71 & -00:00:47.8	&	72.7	&	34.7	&	$12.3^a$	&	0.61 \\
NGC 1365	&	Sb	&	03:33:36.37 & -36:08:25.5	&	23.4	&	62.6	&	$16.98^b$	&	3.05 \\
NGC 1566	&	SABb&	04:20:00.42	& -54:56:16.1	&	44.2	&	49.1	&	$6.61^c$	&	1.30 \\
NGC 2835	&	Sc	&	09:17:52.91	& -22:21:16.8	&	1.3 	&	56.2	&	$8.75^b$	&	1.51 \\
NGC 2997	&	SABc&	09:45:38.79	& -31:11:27.9	&	98.9	&	53.7	&	$11.3^a$	&	1.68 \\
NGC 5068	&	Sc	&	13:18:54.81	& -21:02:20.8	&	152 	&	27.3	&	$6.70^d$	&	1.75 \\
NGC 5236	&	Sc	&	13:39:55.96	& -29:51:55.5	&	54.0	&	15.3	&	$4.66^c$	&	2.93 \\
NGC 7793	&	Scd	&	23:57:49.83	& -32:35:27.7	&	94.4	&	63.5	&	$3.60^c$	&	2.21 \\
\hline
\end{tabular}
    \caption{Astrometric and morphological properties of the eight galaxies investigated in this study. Ra, Dec are taken from the NASA/IPAC Extragalactic Database. Morphological classification, position angles, and inclination data are taken from HyperLEDA. Effective radii are taken to be the half-light radii in the B-band as determined by the Carnegie-Irvine Galaxy Survey \citep{Ho+11}. References for distances: (a) \citet{Cosmicflows1}; (b) \citet{Cosmicflows3}; (c) \citet{Cosmicflows2}; (d) \citet{NearbyGalaxyCatalog}. 
    }
    \label{tab:obs_table}
\end{table*}

In this paper, we use data from the TYPHOON survey (Seibert et al. in prep.). 
Below we briefly summarise the process used to generate these data products -- for further details, see \citet{Poetrodjojo+19}.

Based on observations with the 2.5m du Pont Telescope at Las Campanas Observatory, datacubes were constructed for a sample of large, nearby spiral galaxies with the Progressive Integral Step Method (PrISM). The process builds a 3D datacube (two spatial and one spectral dimensions) through a series of long slit observations for each galaxy in a scanning mode, i.e. shifting each time the long-slit orthogonally to its extension by a constant angular offset. Specifically, a long slit of size $18'\times1.65''$ was used, with native angular resolution along the slit of $0.484''$ (rebinned to $1.65''$ to construct the datacubes), and spectral resolution $R\approx 800$ over the wavelength range $3650-8150$Å. For each galaxy considered in this paper, 200 long-slit observations have been acquired, creating a datacube covering approximately $650\times 200$ pixels. 

Targets were chosen from parent sample of 11HUGS and Local Volume Legacy Survey galaxies \citep{Dale+09}. Galaxies were selected based on their declination ($\leq +10^\circ$), angular diameter ($<18'$, chosen to correspond to the length of the slit), inclination (face-on), and surface brightness (brighter than $21.1$ AB mag/arcsec$^2$ in the B-band).

This work focuses on the local bright galaxy NGC 5236, also known as M83. Lying at a distance of 4.66 Mpc \citep{Cosmicflows2} and with a stellar mass of $10^{10.55} M_\odot$ \citep{Bresolin+16} and a star formation rate of $4.2 M_\odot$ year$^{-1}$ \citep{Leroy+19}, it is one of the closest face-on grand-design spiral galaxies, and is therefore a popular subject for studies on internal metallicity structure \citep[e.g.][]{Bresolin+02, Bresolin+09, Bresolin+16, Boettcher+17, Poetrodjojo+19}.

To ensure our results are valid for a variety of local spiral galaxies, we supplement our study with a sample of seven other galaxies from the TYPHOON/PrISM survey for which the emission line maps have been constructed at time of writing, and the number of identifiable, resolvable \Hii regions is $>300$. Summary properties of these galaxies are listed in Table \ref{tab:obs_table}, and results for these galaxies are described briefly in Appendix \ref{ap:other_galaxies}. The median resolution of each spaxel in this supplementary sample is $68.4$pc, which is more than sufficient for the analysis of small-scale metallicity fluctuations using the geostatistical framework introduced in \GeoGalsI. 

For NGC 5236 (and the other seven supplementary galaxies), the strengths of eight different emission lines and their errors were computed for each spaxel using \texttt{LZIFU}. We imposed several data quality cuts to ensure that the signal-to-noise ratio, S/N, is greater than 3 for all relevant lines, except [O \textsc{ii}]$\lambda\lambda3627,29$, for which we only imposed that S/N > 1.\footnote{At the wavelength of this emission line, the du Pont Telescope is not very sensitive. Imposing a S/N cut of 3 for this line was found to exclude practically all of the data.} Dust extinction was corrected using the model of \citet{ccm89}, before DIG-dominated regions were separated from \Hii regions using three different diagnostics detailed in \ref{ssec:DIG-diagnostics}, yielding three maps of \Hii regions and DIG-dominated regions with high-quality spectral data per galaxy. For each of these maps, the metallicity was determined for each spaxel using four diagnostics: \NSH\ \citep{Dopita+16}, \NO\ \citep{Dopita+13}, \ON\ \citep{Curti+17}, and \RS\ \citep{Curti+20}. Details on the strengths, weaknesses, and calibration process behind each of these diagnostics are given in Appendix \ref{ssec:Z-diagnostics}.

\section{Geostatistical Methods} \label{sec:maths}

In \GeoGalsI, a framework for geostatistical hierarchical modelling was introduced. Furthermore, the semivariogram was discussed as a tool for geostatistical analysis. In this paper, we extend on these ideas, and introduce the concept of universal kriging as a technique that can be used to interpolate metallicities in DIG-dominated spaxels from \Hii region metallicity maps.

We briefly review the \GeoGalsI\ hierarchical geostatistical metallicity model here. The observed metallicity $Z_{\textrm{obs}}$ measured at each location $\vec{x}$ is taken to be a combination of the true metallicity at that location, $Z(\vec x)$, plus some measurement error $\epsilon(\vec x)$:

\begin{equation}
Z_{\rm obs}(\vec x) = Z(\vec x) + \epsilon(\vec x).
\end{equation}

The measurement error is modelled as a stationary Gaussian process with zero mean: $E\left( \epsilon(\vec x) \right) = 0$. It can be completely described by the covariance between any two pairs of data points,  $\textrm{Cov}\left( \epsilon(\vec x), \epsilon(\vec y) \right) $, which in turn depends on the uncertainty of each metallicity measurement, and how spatially correlated measurement errors are expected to be. The measurement error of each emission line for each spaxel is computed for the TYPHOON/PrISM survey using the \texttt{idl} package \texttt{LZIFU}.\footnote{\texttt{LZIFU} is an IFS data fitting tool that uses a penalised pixel-fitting method (pPXF; \citealt{Cappellari+Emsellem04, Cappellari17}) to model stellar continuum and Gaussian components using the Levenberg-Marquardt least-squares method to fit emission lines. The stellar continuum is modelled using the MIUSCAT simple stellar population models \citep{Vazdekis+12} with 13 ages and 3 metallicities. For further details, see \citet{Ho+16}.}

For CCD data, the dominant sources of noise come from dark current, read noise, and shot noise arising from Poisson processes associated with the generation of the detected photons. All these sources will generate white noise that is not spatially correlated between CCD pixels. Furthermore, given the atmospheric conditions at Las Campanas Observatory, the median seeing is $0.6''$ \citep{Persson+90}, which is several times smaller than the size of each spaxel ( $1.65''$). Therefore, the signal itself is also uncorrelated between spaxels.  
Because of this, we are justified in modelling all errors in TYPHOON data as being uncorrelated between spaxels. 

The true metallicity distribution $Z(\vec x)$ is modelled as a nonrandom \emph{process mean} $\mu(\vec x)$, and a spatially varying random component $\eta(\vec x)$ with zero mean:

\begin{equation}
\label{eq:breakdown_true_metallicity}
Z(\vec x) = \mu(\vec x) + \eta(\vec x).
\end{equation}

In this paper, as in \GeoGalsI, $\mu(\vec x)$ is treated as a basic metallicity gradient -- that is,
\begin{equation}
    \mu(\vec{x}) = Z_c +  \gradZ \cdot r(\vec{x}),
    \label{eq:z_grad}
\end{equation}
where $r(\vec{x})$ is the distance between a galaxy's centre and a given location $\vec{x}$. 

The purpose of the function $\eta(\vec{x})$ is to capture the idea that (i) the true metallicities of spaxels are not expected to lie exactly on the mean radial metallicity trend, and (ii) nearby spaxels should have metallicity values that positively correlate with each other.
In \GeoGalsI, $\eta(\vec{x})$ was assumed to take a form such that its covariance matrix matched the five-parameter model developed by \citet{KT18}. In this paper, we use a simpler, more general model for $\eta(\vec{x})$. We model the covariance of the metallicity between two data points separated by a distance of $h=|\vec{x} - \vec{y}|$ using an exponential function:

\begin{equation}
    \text{Cov}(Z(\vec{x}), Z(\vec{y})) = \sigma^2 \exp{\left( - \frac{h}{\phi} \right)}.
    \label{eq:exp_cov}
\end{equation}

This equation has two free parameters: $\sigma^2$, which describes the intrinsic variance within the data caused by random fluctuations around a mean trend (not to be confused with measurement error); and $\phi$, which sets the range over which local variations in metallicity remain correlated. Such a model was chosen for two reasons: first, it is simple, with only two parameters, both of which have physical interpretations in the context of gas-phase physics; yet it remains general enough to capture the overall behaviour of the local metallicity fluctuations. Second, this function is well-studied in the geostatistical literature. It is a special case of the Mátern family \citep{Matern60}, and is known to be positive definite, which is a necessary condition to ensure the covariance matrix between any set of observed data points has no negative entries. We encourage other researchers familiar with this field to construct their own covariance functions using models of ISM turbulence\footnote{For a list of known positive-definite functions that can be applied in order to model spatial covariance, see e.g. \citet{Gneiting97}.}.

The semivariogram, among other applications, can be applied to distinguish the small scale metallicity fluctuations, $\eta$, from measurement error, $\epsilon$, as discussed in \GeoGalsI. Formally, the semivariogram $\gamma(h)$ is defined to be half the variance between data points separated by a distance of $h$. When the covariance function depends only on the distance between data points (that is, there exists a $C(r)$ such that if $\|\vec{x} - \vec{y} \|=r$ then $\text{Cov}(Z(\vec{x}), Z(\vec{y}))=C(r)$), the semivariogram can be calculated using the formula $\gamma(h) = C(0) - C(h)$. Because of this, Equation \ref{eq:exp_cov} induces the following form for the semivariogram:

\begin{equation}
    \gamma(h) = \sigma^2 \left(1- \exp{\left( -h/\phi \right)} \right).
    \label{eq:theoretic_svg}
\end{equation}

For spatially correlated data for which there is no spatially-correlated measurement error, $\gamma(0)=0$. As $h$ increases, $\gamma(h)$ increases, until the data are so separated that they can be considered uncorrelated, at which point $\gamma(h)$ approaches the intrinsic variance of the correlated data.
For this model, $\sigma^2$ sets the height of the semivariogram, and $\phi$ determines the physical scale $h$ at which it flattens out. Additional uncorrelated noise (e.g. from measurement error) adds a constant factor to the measured semivariogram at all separations, and can thus be identified by computing $\gamma(0)$. Further details on the semivariogram and its role in geostatistical data analysis can be found in \citet{Wikle+19}.

\subsection{Fitting parameters for a geostatistical hierarchical model}
\label{ssec:fitting}

Overall, our model for the 2D internal metallicity structure of the ISM of local galaxies has $4$ parameters; two of which describe the global trend of metallicity with radius $(Z_c, \gradZ)$, and two of which describe the small-scale metallicity fluctuations associated with stochastic star-formation and diffusion of metals through the ISM $(\sigma^2, \phi)$.

In \GeoGalsI, $Z_c$ and $\gradZ$ were fit using a weighted least-squares approach, taking into account only the measurement error $\epsilon(\vec{x})$ of each spaxel, and not accounting for the small-scale deviations of the metallicity around the median value, $\eta(\vec{x})$. In hindsight, a semivariogram analysis showed that these local spiral galaxies are far from well-mixed, and the magnitude of these small-scale fluctuations was of the same order as the size of the measurement errors. Therefore, in order to properly model the metallicity gradient of these galaxies, the size and structure of small-scale metallicity fluctuations must be accounted for.

Such an approach is conceptually similar to the framework employed by \citet{Clark+19} and \citet{Williams+22} (described in e.g. \citealt{Hogg+10}) to fit the mean radial trend of their geostatistical models. This approach employs an additional parameter to describe the intrinsic scatter around the linear trend, and fits this parameter together with the linear fit using a maximum-likelihood method. However, this method only accounts for the additional variance caused by small-scale fluctuations around the mean trend, without accounting for correlations between nearby data points.

In this paper, we demonstrate an improved, statistically robust approach, wherein all four parameters required to build a geostatistical model for a galaxy are fit simultaneously using maximum-likelihood estimation (ML), following the procedure described in \citet{Diggle+Riberio07}. By fitting all of these parameters at the same time, the uncertainty in their estimated values is reduced \citep{MLE}. We detail this method below.

Let $\bm{Z}_{\rm obs}$ be a vector of $N$ observed metallicity values over a spatial domain, and let $\bm{C}_\epsilon$ be the $N \times N$ covariance matrix that completely describes the structure of correlated measurement errors between data points. For the case of TYPHOON galaxies, all measurement errors are modelled as uncorrelated and so $\bm{C}_\epsilon$ is a diagonal matrix, with elements in the diagonal equal to the variance in each metallicity measurement. These metallicity measurement uncertainties are computed from the uncertainty in emission line flux using linear error propagation (described in Section \ref{ssec:Z-diagnostics}; see Equation \ref{eq:Z_error}), and are fixed for each galaxy.

Let $\bm{D}$ be the $N \times 2$ matrix with every element in the first column equal to $1$ and the second column given by the deprojected distance from each observation point to the galaxy centre. Let $\bm \beta = [Z_c, \gradZ]^T$ - where $^T$ is the transpose operator - so that $\bm{D}\bm{\beta}=\bm{\mu}$, the metallicity of each spaxel predicted by the linear metallicity gradient model. 
\footnote{More generally, $\bm{D}$ may be a matrix containing the values of any number of covariates for each data point. Each covariate is a variable that is expected to influence a galaxy's local metallicity, such as the local star formation rate of each spaxel, or whether or not each spaxel is contained within a spiral arm. In this general case, $\bm \beta$ would be a vector containing the parameters describing the global trend, such that $\bm{D}\bm{\beta}$ is still the predicted mean metallicity for all data points, $\bm{\mu}$.}
Let $\bm{C}_Z(\sigma^2, \phi)$ be the covariance matrix associated with local metallicity fluctuations $\eta(\vec{x})$. Because the deprojected distance between each pair of spaxels is known, the value of each entry in this matrix can be calculated using Equation \ref{eq:exp_cov}. 

Using our geostatistical framework, we model the vector $\bm{Z}_{\rm obs}$ as being drawn from a $N$-dimensional Gaussian distribution, with mean given by $\bm{\mu} = \bm{D}\bm{\beta}$, and total covariance matrix given by $\bm{C}_\epsilon + \bm{C}_Z(\sigma^2, \phi)$. The natural logarithm of the likelihood of drawing $\bm{Z}_{\rm obs}$ from this distribution is then an analytic function of our four parameters:

\begin{equation}
    \begin{split}
    \mathcal{L}(\bm{\beta}, \sigma^2, \phi) = -\frac{1}{2} \Big[ N \log(2\pi) + \log \left( \left| \bm{C}_\epsilon + \bm{C}_Z(\sigma^2, \phi) \right| \right) + \\ \left( \bm{Z}_{\rm obs} - \bm{D}\bm{\beta} \right)^T\left( \bm{C}_\epsilon + \bm{C}_Z(\sigma^2, \phi) \right)^{-1} \left( \bm{Z}_{\rm obs} - \bm{D}\bm{\beta} \right) \Big] 
    \end{split}.
    \label{eq:full_LL}
\end{equation}

This function can be maximised using many numerical methods. To simplify the problem and speed up computation, we use a profile log-likelihood approach. For any specified values of $\sigma^2$ and $\phi$, the most likely values for $\bm{\beta}$ are given by the following linear equation:

\begin{equation}
    \hat{\bm{\beta}}(\sigma^2, \phi) = \left(\bm{D}^T \bm{V}(\sigma^2, \phi)^{-1} \bm{D}\right)^{-1} \bm{D}^T \bm{V}(\sigma^2, \phi) ^{-1} \bm{Z}_{\rm obs},
    \label{eq:best_beta}
\end{equation}
where $\bm{V}(\sigma^2, \phi) = \bm{C}_\epsilon + \bm{C}_Z(\sigma^2, \phi)$. This equation can be substituted back into Equation \ref{eq:full_LL}, in order to give a function of $\sigma^2$ and $\phi$ that, when maximised, gives their maximum likelihood estimates. This approach effectively reduces the complexity of this task from a four-dimensional optimisation problem into a two-dimensional one.

The uncertainty in the best-fitting parameters for a kriging model can be estimated in many ways, including using robust empirical methods such as bootstrapping, or by taking a Bayesian approach and analysing the posterior distribution of the parameters. We estimate the uncertainty of $\sigma^2$ and $\phi$ by computing the inverse of the Hessian matrix, which for exponentially correlated data provides a good approximation of the covariance matrix for these parameters \citep{MLE}. To find the uncertainty in $Z_C$ and $\gradZ$, we use the formula for the covariance matrix for a generalised least-squares fit, using the maximum likelihood estimates of $\sigma^2$ and $\phi$:

\begin{equation}
    \text{Cov} \left( \hat{\bm{\beta}} \right) = \left(\bm{D}^T \bm{V}(\sigma^2, \phi)^{-1} \bm{D}\right)^{-1}.
    \label{eq:unc_best_beta}
\end{equation}

\subsection{Universal Kriging}
\label{ssec:kriging}
When the structure of spatial correlations is known for a process, measured data points can be used to predict nearby unknown values. \emph{Universal kriging} is a method by which the true metallicity can be estimated at any point within a galaxy, accounting for large scale trends using a Gaussian additive model fit via a generalised least squares regression (GLS), and modelling small-scale variation using a known covariance function \citep{Matheron69, UK}.

To define universal kriging, we follow \citet{Wikle+19}. Using the same notation as in Section \ref{ssec:fitting}, our observations $\bm{Z}_{\rm obs}$, their errors $\bm{C}_\epsilon$, and the best-fitting geostatistical model may be used to estimate $Z$ at any unknown data point, $\vec{x_0}$. Let $\bm{c_0}$ be the vector with $i$th element given by $\text{Cov}(Z(\vec{x_0}), Z(\vec{x_i}))$ - that is, $\bm{c}_0$ gives the modelled covariance between this vector and each observed data point. Finally, let $\bm{d}_0$ be the valuation of each covariate function at this point -- for the simple metallicity gradient model, $\bm{d}_0 = [1, r(\vec{x_0})]^T$. Then we may estimate the value of $Z$ at this point and the uncertainty of this estimate using the following pair of equations:

\begin{eqnarray}
\hat Z(\vec{x_0}) &=& \bm{d}_0 {\boldsymbol{\beta}}+\mathbf{c}_{0}^{T} \bm{V}(\sigma^2, \phi)^{-1}\left(\mathbf{Z}-\mathbf{D} {\boldsymbol{\beta}}\right), \label{eq:krig_predict}\\
\sigma_{Z,uk}^2 &=& \sigma^2 - \mathbf{c}_{0}^{T} \bm{V}(\sigma^2, \phi)^{-1} \mathbf{c}_{0} + k.
\label{eq:krig_error}
\end{eqnarray}

Here, $\mathbf{c}_{0}^{T} \bm{V}(\sigma^2, \phi)^{-1} \mathbf{c}_{0}$ represents a factor by which the variance associated with this data point is reduced after taking into account its correlation with known (but uncertain) data points, 
and $k$ is the additional uncertainty that comes about from estimating $\bm{\beta}$ values using the GLS method:

\begin{equation}
    \begin{split}
k \equiv\left(\bm{d}_0 -\mathbf{D}^{T} \bm{V}(\sigma^2, \phi)^{-1} \mathbf{c}_{0}\right)^{T}\left(\mathbf{D}^{T} \bm{V}(\sigma^2, \phi)^{-1} \mathbf{D}\right)^{-1} \\ \left(\bm{d}_0-\mathbf{D}^{T} \bm{V}(\sigma^2, \phi)^{-1} \mathbf{c}_{0}\right).
    \end{split}
\end{equation}

Because this mathematics is powered by linear algebra, it is simple to extend these definitions to work for multiple points at a time (see, e.g. \citealt{Cressie93}). This allows metallicity estimates to be computed rapidly for a grid of data points, leading to highly detailed maps of the predicted internal metallicity structure of galaxies, or for a list of data points where the metallicity cannot be measured directly, such as DIG-dominated spaxels. In this paper, we will show examples of both of these constructions.

\section{Results}
\label{sec:results}

\begin{figure*}
    \centering
    \includegraphics[width=\textwidth]{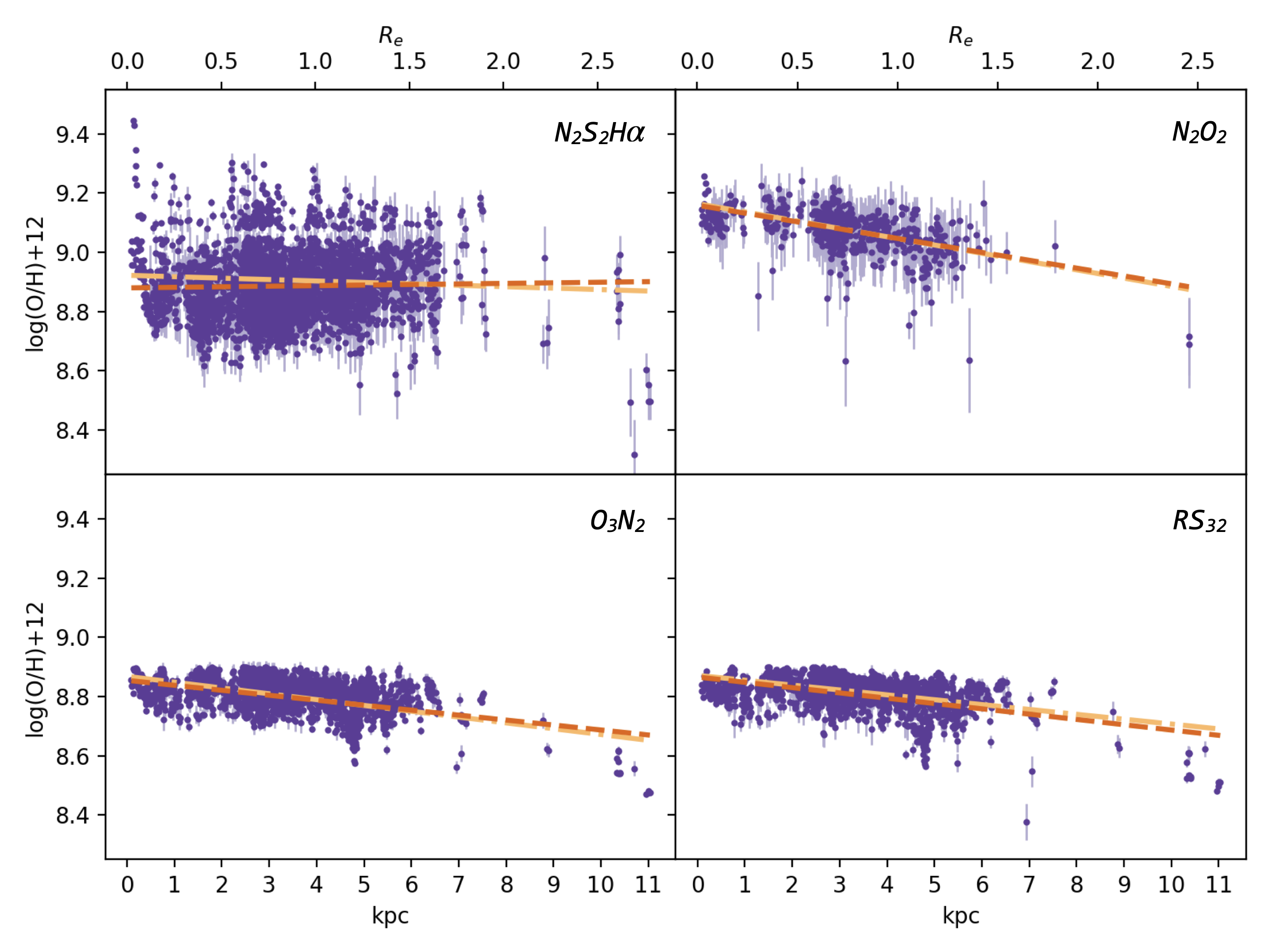}
    \caption{Radial metallicity profiles of \Hii regions in the galaxy NGC 5236, computed using the 4 different metallicity diagnostics described in Section \ref{ssec:Z-diagnostics}. Uncertainties in the metallicity of each \Hii spaxel are computed using linear error propagation, and are shown as error bars. For each diagnostic, the best fitting metallicity gradient model computed using a ML method accounting for small-scale correlations is shown with an orange dashed line. These are consistent with the best fitting metallicity gradients computing using the WLS method, shown as the golden dash-dotted line.}
    \label{fig:gradients}
\end{figure*}

\subsection{The geostatistical model for H\textsc{II} regions}
\label{ssec:M83_model}

In this Section, we present the results of geostatistical model fitting for NGC 5236. Using the [S \textsc{ii}]/H$\alpha$ diagnostic of \citet{Kaplan+16} described in Section \ref{ssec:S2-diagnostic}, $3664$ \Hii spaxels and $19607$ spaxels with significant DIG contamination were identified. The number of spaxels with metallicities determined after imposing S/N cuts for the emission lines used in each diagnostic are listed in Table \ref{tab:n_spaxels}.
Using the \Hii regions, the four free parameters of the geostatistical model described in Section \ref{sec:maths} were fit, using the ML method outlined in Section \ref{ssec:fitting}. 

\begin{table}
    \centering
    \begin{tabular}{l|r|r}
    \hline
    Diagnostic & \# of \Hii regions & \# of DIG regions \\
    \hline
    \begin{tabular}{@{}l@{}}\NSH \\ \citep{Dopita+16}\end{tabular} & 3664 & 19602 \\
    \begin{tabular}{@{}l@{}}\NO \\ \citep{Dopita+13}\end{tabular} & 623 & 629 \\
    \begin{tabular}{@{}l@{}}\ON \\ \citep{Curti+20}\end{tabular} & 2459 & 6625 \\
    \begin{tabular}{@{}l@{}}\RS \\ \citep{Curti+20}\end{tabular} & 2482 & 6628 \\
    \hline
    \end{tabular}
    \caption{Number of \Hii spaxels and spaxels with significant DIG contamination in NGC 5236 for which metallicities have been determined with each diagnostic. Here, DIG-dominated regions are separated from \Hii regions using the methodology of \citet{Kaplan+16} described in Appendix \ref{ssec:S2-diagnostic}. Further details on the metallicity diagnostics presented here are provided in Appendix \ref{ssec:Z-diagnostics}.}
    \label{tab:n_spaxels}
\end{table}

In Figure \ref{fig:gradients}, we show the most likely large-scale metallicity trend, $\mu(\vec{x})$, according to this model, for NGC 5236. All \Hii spaxels with metallicities determined for each diagnostic are shown, with error bars reflecting the uncertainty in the metallicity obtained by linear error propagation. For \ON, \RS, and \NSH, the error associated with each spaxel is small, with median values of $0.014,$ $0.013,$ and $0.025$ dex, respectively. Conversely, for \NO the median error associated with each spaxel is $0.058$ dex. This large uncertainty stems from the lack of sensitivity of the Wide Field CCD instrument on the du Pont telescope at the wavelength of the [O \textsc{ii}]$\lambda\lambda3726,29$ emission line, and our subsequent decision to use a less stringent S/N cut for this line.

\begin{figure*}
    \centering
    \includegraphics[width=\textwidth]{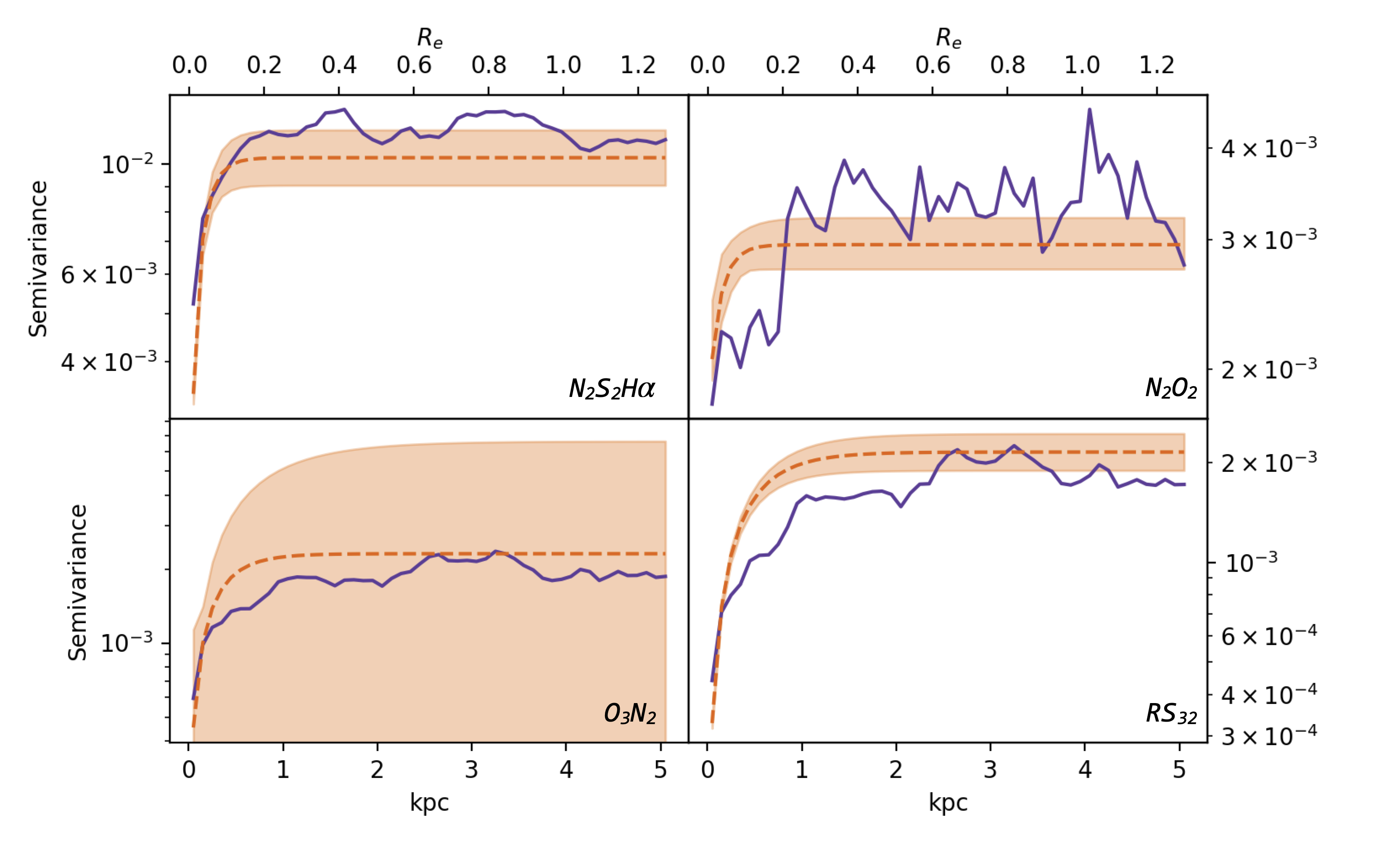}
    \caption{Semivariograms showing the small-scale metallicity structure of NGC 5236, using a suite of metallicity diagnostics. Purple solid lines show the empirical semivariogram. Orange dashed lines show the semivariogram predicted from the model of best fit. Orange shaded regions show the $1\sigma$ error in the fitted curve (errors on the semivariance estimation from the data at each separation shown are negligible). In all cases, the behaviour of the model shows good agreement with the data, although for the \ON\ diagnostic, the uncertainties on the parameters for the most likely model for small-scale structure are very large.}
    \label{fig:semivariograms}
\end{figure*}


\begin{table*}
    \centering
    \begin{tabular}{l|c|c|r|r|c|c}
      \hline
       Diagnostic & Fitting method & $Z_c$ & $\gradZ$ (dex kpc$^{-1}$) & $\gradZ$ (dex $R_e^{-1}$) & $\sigma^2$ & $\phi$ (kpc) \\
       \hline
      \NSH & ML model & 8.88$\pm$0.012 & 0.0019$\pm0.012$ & 0.0076$\pm 0.012$ &0.0100$\pm0.0013$&$0.1329 \pm 0.015$\\
                    & WLS      & 8.92$\pm$0.004  &-0.0048$\pm 0.001$ &-0.0192$\pm0.005$ &-&-\\
    \hline
      \NO  & ML model & 9.16$\pm$0.009 &-0.0265$\pm0.009$ &-0.1053$\pm0.010$ &0.0013$\pm0.0002$&0.1331 $\pm 0.091$\\
                    & WLS      & 9.16$\pm$0.005 &-0.0279$\pm0.002$ &-0.1110$\pm0.006$ &-&-\\
    \hline
      \ON  & ML model & 8.85$\pm$0.009 &-0.0168$\pm0.009$ &-0.0665$\pm0.008$ &0.0022$\pm0.0024$&0.2872 $\pm 0.386$\\
                    & WLS      & 8.87$\pm$0.002 &-0.0196$\pm0.007$ &-0.0780$\pm$0.003 &-&-\\
    \hline
      \RS  & ML model & 8.86$\pm$0.011 &-0.0179$\pm0.011$ &-0.0712$\pm0.009$ &0.0021$\pm0.0003$&0.3980 $\pm 0.053$\\
                    & WLS      & 8.91$\pm$0.002 &-0.0106$\pm0.006$  &-0.0420$\pm0.002$ &-&-\\
      \hline
    \end{tabular}
    \caption{Results of ML model fitting for the geostatistical model, compared to a WLS approach to compute the metallicity gradient using each metallicity diagnostic for NGC 5236.}
    \label{tab:Z_grads}
\end{table*}

\begin{figure*}
    \centering
    \includegraphics[width=0.99\textwidth]{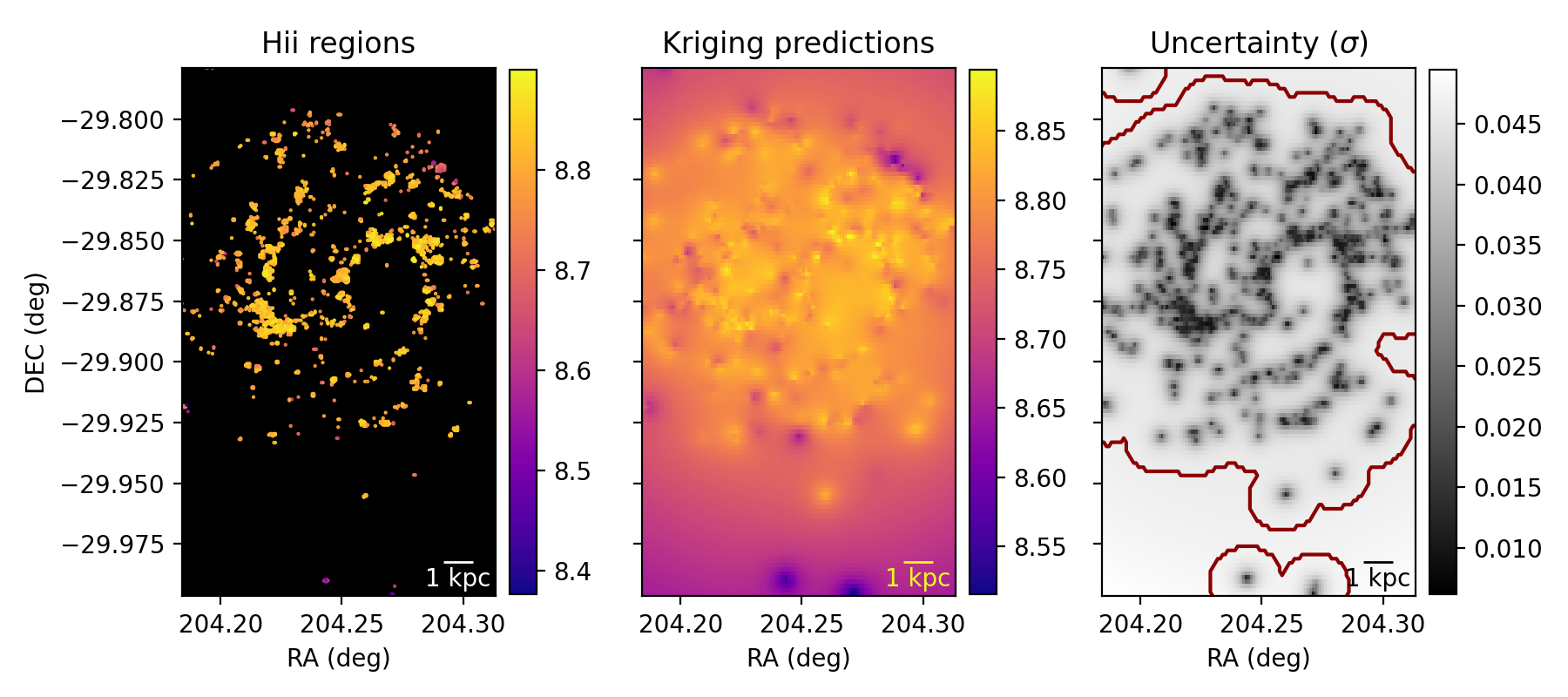}
    \caption{\emph{Left:} Measured metallicity values, in units of $\log([$O/H$])+12$, for H\textsc{ii} regions in NGC 5236, determined using the \RS\ diagnostic. \emph{Centre:} An interpolated metallicity map, computed using the technique of universal kriging, modelling small-scale metallicity fluctuations with a Gaussian model with an exponential correlation function. \emph{Right:} Uncertainties (standard deviation) of the kriging predictions at each location. At distances greater than $3\phi$ from any \Hii spaxel (red contour), the uncertainty of the metallicity estimated for any point is approximately the uncertainty given by a metallicity gradient model.}
    \label{fig:krig}
\end{figure*}

The dearth in metallicities computed with the \NSH\ diagnostic between $\log (O/H) + 12 = 9.05$ and $9.1$ stems from the fact that this diagnostic contains a correction term, for metallicities greater than 9.05. This introduces a discontinuity into the metallicity diagnostic that can be seen in Figure \ref{fig:gradients}. The uncertainties due to this correction factor are completely accounted for in this analysis, as described in Appendix \ref{ssec:Z-diagnostics}.

In \GeoGalsI,
local star-forming galaxies were found to have significant variations in metallicity on small-scales due to inefficient mixing.
The geostatistical method we present in this work accounts for these small-scale metallicity fluctuations when fitting the global trend in a self-consistent way, by simultaneously modelling the local covariance structure and using maximum-likelihood estimation. In Figure \ref{fig:gradients}, we compare the metallicity gradient models obtained using this method to the metallicity gradients one would obtain using a simple weighted least-squares (WLS) approach. The WLS method assumes (i) that there is no correlated variance between local data points beyond what is described by the metallicity gradient, (ii) that all data points should, in reality, lie exactly on the line-of-best-fit found by this method, and (iii) that any deviations in the metallicity of individual spaxels about this line is caused by measurement error \citep{Hogg+10}. 
For this galaxy, the radial metallicity profiles determined using this methodology agree with the popular WLS method for all diagnostics except \NSH, where the geostatistical ML method favours a slightly positive metallicity gradient, as compared to the slightly negative metallicity gradient that is determined with WLS. However, in this case both models agree that the metallicity gradient is very close to $0$, and are consistent within one standard deviation. 

The metallicity gradients returned by each of these methods for this galaxy under each diagnostic are listed in Table \ref{tab:Z_grads}. We note that the uncertainties in the values of $Z_c$ and $\gradZ$ computed using the ML method are larger than those determined using the WLS method, which seemingly implies that the WLS method produces more precise metallicity gradient estimates. We stress that this is not the case. The smaller uncertainty in the WLS fitted parameters come from the erroneous assumption that there is no scatter about the radial metallicity trend besides measurement error, which would imply that the data points are very reliable tracers of the metallicity gradient, and only a small number of data points are needed in order to recover this global trend. On the other hand, under the geostatistical model, data are not expected to perfectly trace the average radial trend, leading to more realistic uncertainties in the recovered metallicity gradient parameters. 

In Figure \ref{fig:semivariograms}, the small-scale structure of the metallicity fluctuations determined under each diagnostic are illustrated with semivariograms (see \GeoGalsI\ for a thorough explanation of semivariograms and their role in extragalactic astronomy). Solid purple lines show the empirical semivariograms, computed using bins of width $0.1$ kpc, up to a separation of $5$ kpc. With all diagnostics, significant structure is seen in these semivariograms, revealing correlations between spaxels separated by up to 1-2 kpc. Similar behaviour has also been seen for a population of local star-forming galaxies observed by the PHANGS collaboration (\citealt{Kreckel+20}, \GeoGalsI). For \NSH, \ON, and \RS, the true variance about the mean metallicity at each radius ($\sigma^2$ in our model) is $\sim 1$ order of magnitude larger than the median variance in metallicity caused by error measurements. For \ON, \NO, and \RS, the height of these semivariograms is broadly consistent with the height of the semivariograms computed using PHANGS data in \GeoGalsI. For \NSH, the height of the semivariogram is much larger. This can be understood by examining Figure \ref{fig:gradients}: under this diagnostic, there is a much larger scatter around the mean metallicity at any radius than there is when any other diagnostic is used.

\begin{figure*}
    \centering
    \includegraphics[width=0.99\textwidth]{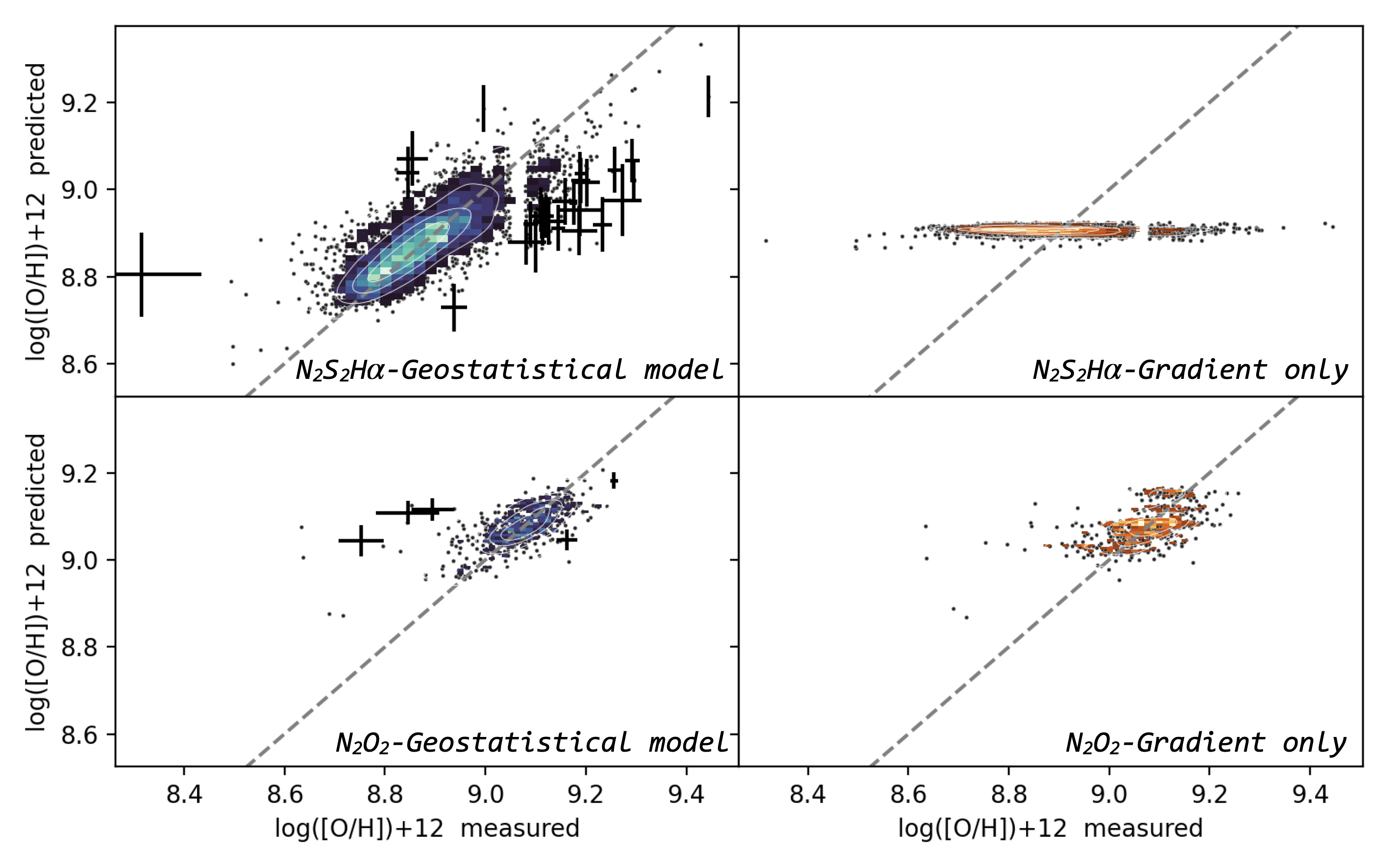}
    \caption{Scatter plots comparing the metallicities predicted for \Hii regions using 10-fold cross validation to the true measured metallicities of these points. \emph{Left:} Comparing metallicities predicted with universal kriging to the measured metallicity of each spaxel, computed using the \NSH\ (top) and \NO\ (bottom) diagnostics. 3$\sigma$ outliers are shown with error bars. \emph{Right:} The same, but using predictions from a simple metallicity gradient model computed using weighted least-squares, without accounting for small-scale metallicity fluctuations. As this method does not return the uncertainty associated with each metallicity prediction, outliers are not explicitly shown.}
    \label{fig:CV_scatter}
\end{figure*}

\begin{figure*}
    \ContinuedFloat
    \centering
    \includegraphics[width=0.99\textwidth]{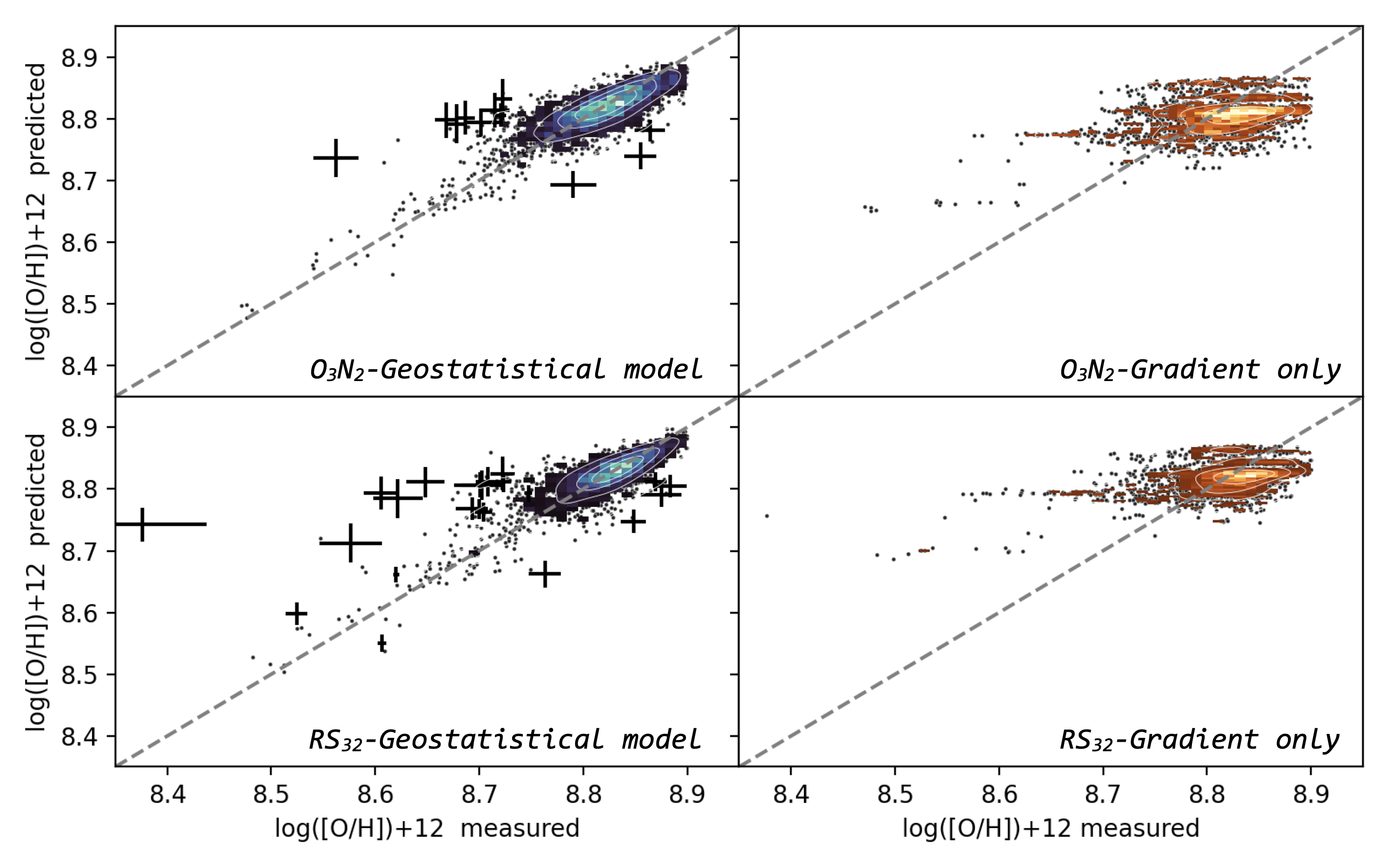}
    \caption{ -- \textit{continued}. Comparison of cross-validation results for the \ON\ (\emph{top}) and \RS\ (\emph{bottom}) diagnostics, using the full geostatistical model (\emph{left}) with 3-$\sigma$ outliers shown with error bars, and a simple metallicity gradient (\emph{right}).}
\end{figure*}

Dashed orange lines show the semivariograms associated with the most likely exponential model of $\eta(\vec{x})$ for each diagnostic, with maximum likelihood parameters listed in Table \ref{tab:Z_grads}. For all diagnostics, this model adequately captures the most important features of the small-scale metallicity structure; namely, the spatial extent of these local metallicity fluctuations ($\phi$), and the total amount of correlated variance ($\sigma^2$). However, improvements could be made in modelling the general shape of this correlation function, as correlations between nearby datapoints seem to decrease with distance less rapidly than is predicted by an exponential covariance function. To properly model this behaviour, a physically motivated positive-definite covariance function may need to be constructed, using theoretical ideas about turbulence in the ISMs of local galaxies (in the style of e.g. \citealt{KT18}). 
In this respect, semivariograms such as those constructed for this galaxy may be used as a tool to assist theoretical astronomers in the development of testable models of local ISM enrichment and metal transport.

Shaded regions in Figure \ref{fig:semivariograms} show the 1-$\sigma$ uncertainty on the theoretical semivariogram produced by the fitted parameters, computed using Equation \ref{eq:unc_best_beta}.\footnote{The uncertainties in the estimate of the measured semivariance at each separation (purple line) is very small, and is not shown. This is because the uncertainty in the estimate of the variance at each separation is proportional to the number of pairs of data points separated by that distance, and this number is very large ($>990$ for\NO and $>9700$ for all other diagnostics) at all separations shown.} For the model based on the \ON\ diagnostic, the uncertainties in the best fit values of $\phi$ and $\sigma^2$ reported in Table \ref{tab:Z_grads} are very high. While the maximum likelihood geostatistical model for the metallicity distribution under this diagnostic is broadly consistent with the models obtained using other diagnostics, the size of the uncertainty in the fit implies that this data is also consistent with no additional uncertainty. We can understand the size of these parameter uncertainties by looking at the semivariogram associated with this data, shown in Figure \ref{fig:semivariograms}. Clearly, the data do exhibit spatial correlation. However, it is difficult to determine both the range over which data become uncorrelated and the limiting value of the semivariance as the separation tends towards infinity. This indicates that the exponential correlation model, chosen for its simplicity, may not be sophisticated enough to capture the small-scale metallicity trends associated with this diagnostic. A similarly high uncertainty is present in the ML estimate for $\phi$ when the \NO\ diagnostic is used, which further hints that the simple exponential model with a single correlation scale for $\eta$ may not be sufficient for capturing all of the details of the metallicity structure of the ISM.

Using the geostatistical models calibrated to \Hii region data, predictions of the metallicity can be made at unmeasured locations using the technique of universal kriging, described in Section \ref{ssec:kriging}. We illustrate this process using the \RS\ diagnostic for NGC 5236 in Figure \ref{fig:krig}. This diagnostic was chosen for visualisation because it contains both a significant metallicity gradient, and substantial small-scale metallicity fluctuations. Using the World Coordinate System (WCS) data of the PrISM datacube for this galaxy, a grid of 128$\times$128 sky locations was constructed, spanning the FOV of the constructed metallicity map. At each of these locations, the metallicity was predicted using Equation \ref{eq:krig_predict}, with uncertainty given by Equation \ref{eq:krig_error}. 

In the left-hand panel of Figure \ref{fig:krig}, we show the metallicity measurements for all \Hii spaxels. Together with our geostatistical model, these metallicities and their uncertainties are used to predict the metallicity at each location in NGC 5236. In the middle panel, we show the predicted metallicity map. Within this predicted map, metallicity decreases as distance from the galactic centre decreases -- but not smoothly or uniformly. Instead, small-scale stochastic metallicity fluctuations can be seen, with a characteristic scale of $\phi \approxeq 400$ pc, revealing locations with local chemical enrichment and depletion.
Interestingly, there appears to be some enrichment along the main spiral arms of NGC 5236 in the predicted 2D metallicity distribution in concordance with the chemical carousel model of \citet{Ho+17}, even though such enrichment along spiral arms was not modelled explicitly. A quantitative exploration of this effect with a wider sample of galaxies, using geostatistical methods to improve completeness of data in the inter-arm regions of spiral galaxies, is an avenue for future research.

In the right-hand panel of Figure \ref{fig:krig}, the uncertainty associated with each metallicity prediction is plotted. When predictions are made close to an observation point, the metallicity is expected to be highly correlated with the metallicity measured at the \Hii region, and so the uncertainty of the prediction is reduced to $\sim 0.01$ dex, similar to the uncertainty in metallicity measurements at that datapoint. As the distance from a \Hii region increases, the uncertainty likewise increases, in a manner that is consistent with the semivariogram for this diagnostic shown in Figure \ref{fig:semivariograms}. When the distance from any observed data point is larger than $\sim 1.2$ kpc ($\sim 3\phi$, shown by the red contour in the Figure), the error in the predicted metallicity is equal to the true scatter around the metallicity gradient, plus the uncertainty associated with the metallicity gradient computed (see Equation \ref{eq:krig_error}). 

\subsection{Validation}
\label{ssec:validation}

To verify the accuracy of this model, we predict the metallicity of each \Hii region using a model trained on a subset of the data. We then compare metallicities predicted by the geostatistical model to those measured directly using strong emission line diagnostics.

Formally, this was done by performing a 10-fold cross-validation \citep{James+13}. Each data point was randomly assigned to one of $10$ groups. For each of these groups, a geostatistical model of the form described in Section \ref{sec:maths} was fit using the ML method detailed in Section \ref{ssec:fitting}, calibrated using only the \Hii spaxels in the other $9$ groups (the \textit{training set}). Then, the value of the metallicity at each spaxel in the selected group (the \textit{testing set}) was estimated using universal kriging, combining knowledge of the large-scale metallicity trend, $\mu(\vec{x})$ with correlations to nearby \Hii regions in the training set under our model of local metallicity fluctuations $\eta(\vec{x})$ .

Figure \ref{fig:CV_scatter} shows the results of this analysis. In the left hand panels, 2D histograms show the distribution of metallicities predicted using this method of universal kriging, as compared to the observed metallicities of each spaxel. All 3$\sigma$ outliers are shown with error bars in the predicted and measured metallicity. In the right hand panels, we repeat this cross-validation analysis, but instead predicting the metallicity of each \Hii region using a simple metallicity gradient model computed using WLS, without accounting for any correlations in the metallicities of nearby \Hii regions. Because this method does not produce uncertainties in the predicted metallicities of each data point, we do not show outliers explicitly.

\begin{figure*}
    \centering
    \includegraphics[width=0.8\textwidth]{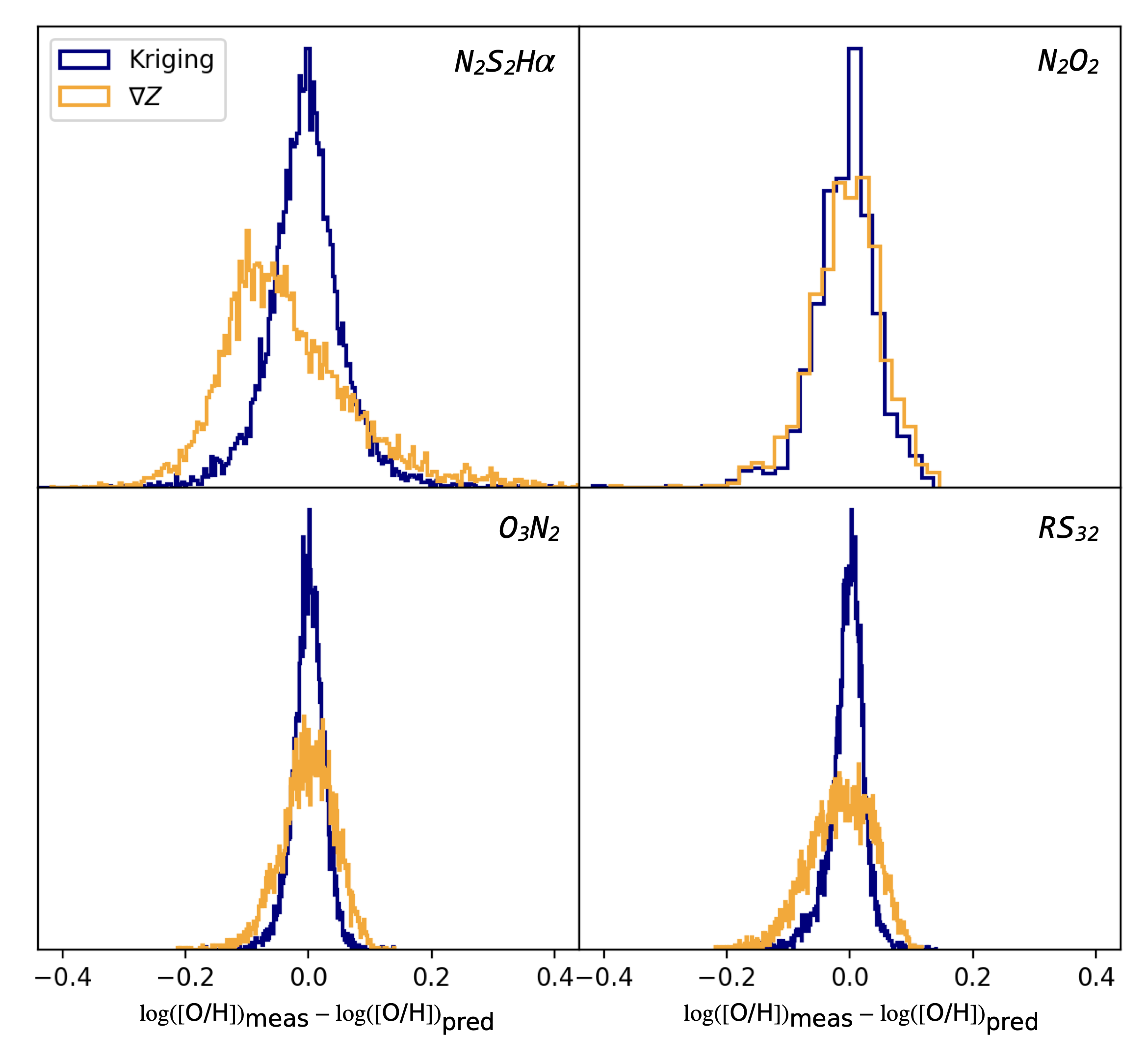}
    \caption{Histograms of the difference between the metallicities measured for each \Hii spaxel $Z_{\textrm{meas}}$, and the predicted metallicities using 10-fold cross validation $Z_{\textrm{pred}}$, using universal kriging (dark blue lines) and a simple metallicity gradient approach (gold lines). For all diagnostics, the metallicities predicted via universal kriging show less bias and are less scattered than predictions that do not account for local spatial trends.}
    \label{fig:CV_hists}
\end{figure*}

Universal kriging shows the greatest improvement over the WLS method for NGC 5236 when the \NSH\ diagnostic is used. Using this diagnostic, the metallicity gradient of this galaxy is very close to zero, and so metallicity prediction methods that rely solely on the metallicity gradient will predict the metallicity of all data points to be approximately equal. Clearly, this is not the case. \NSH\ metallicities of \Hii spaxels in this galaxy have been measured to span a range of \mbox{$\log($[O/H]$)+12=8.6-9.2$}. This large variance in metallicity can, however, be modelled using local metallicity fluctuations. When universal kriging is used, a large majority ($76.20\%$) of predicted metallicities lie within 1 standard deviation of the measured metallicity value, and $95.63\%$ agree to 2$\sigma$. Furthermore, for all diagnostics tested, 
less than $5\%$ of predicted metallicities differ from the true metallicities by more than 2$\sigma$,
and the proportion of $3\sigma$ outliers is less than $1\%$.
These proportions are in line with the theoretical numbers of predictors within $1\sigma$, $2 \sigma$ and $3 \sigma$ expected from the 68-95-99.7$\%$ heuristic, under the Gaussian assumption of the kriging predictor.

For all other diagnostics, the improvement in predictions gained from the use of universal kriging is more mild, but still significant. In particular, when a simple $Z$-gradient model is used, the metallicities of the most metal-rich spaxels is under-predicted, and the metallicities of the most metal-poor spaxels are over-predicted. This demonstrates that universal kriging is a powerful tool for capturing the detailed behaviour of all spaxels in the galaxy, including departures from a linear gradient trend.

To make the improvement in the accuracy of predictions gained by creating a geostatistical model clear, in Figure \ref{fig:CV_hists} we plot histograms 
showing the difference between the metallicities predicted by this model ($Z_{\textrm{pred}}$, measured in units of $\log([$O/H$])+12$) and the metallicities measured ($Z_{\textrm{meas}}$, reported in the same units)
for each of the $4$ studied metallicity diagnostics. Two features are clearly visible from this graph. Firstly, for all diagnostics, the distribution of $Z_{\textrm{meas}} - Z_{\textrm{pred}}$ is narrower for predictions obtained using universal kriging than those obtained using a metallicity gradient (although this improvement is only marginal for \NO), reflecting the increased accuracy of the geostatistical model. Secondly, for \NSH, \ON\ and \RS, $Z_{\textrm{meas}} - Z_{\textrm{pred}}$ is slightly biased for the WLS model, with peak values of $Z_{\textrm{meas}} - Z_{\textrm{pred}}$ of $-0.08, +0.02,$ and $+0.02$ dex, respectively. Conversely, the distributions gained from universal kriging are unbiased. This is because universal kriging is the optimal unbiased linear estimator for Gaussian processes where the global trend is a priori unknown, and correlation between nearby data points cannot be assumed to be negligible \citep{Wikle+19}, whereas the WLS method is only unbiased under the assumption that the only deviation of data points from a linear trend line is due to uncorrelated Gaussian error \citep{Hogg+10}, which is not a valid assumption for this kind of data.

\section{Predicting metallicities in DIG-dominated regions}
\label{sec:vs-dig}

\begin{figure*}
    \centering
    \includegraphics[width=0.99\textwidth]{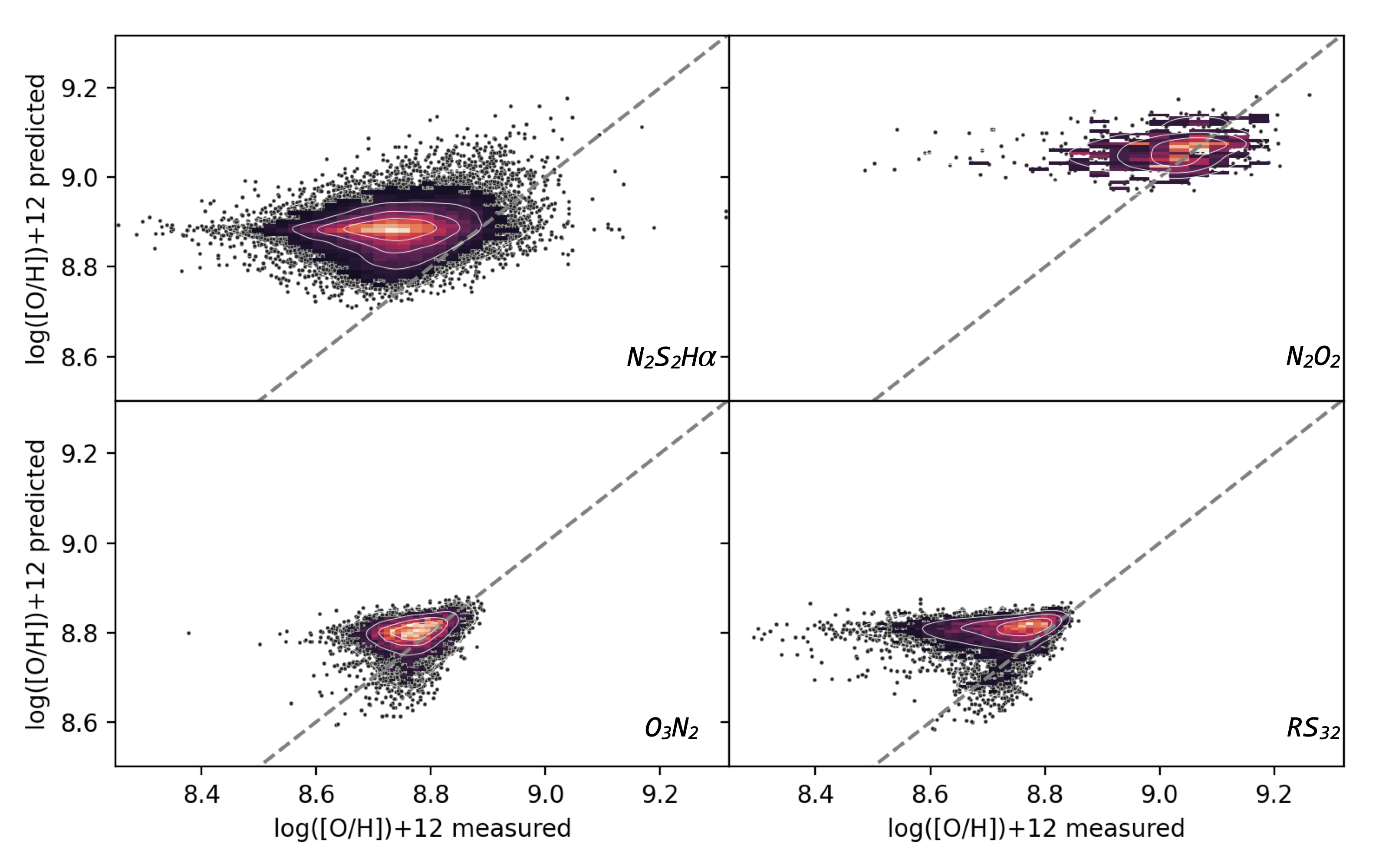}
    \caption{Scatter plots comparing the metallicities predicted for DIG-dominated regions in NGC 5236 using universal kriging, to those measured directly using \Hii region calibrated strong emission-line diagnostics. We find that the \Hii region calibrated diagnostics yield metallicities for the DIG-dominated spaxels that do not agree with our fitted models of metal mixing.}
    \label{fig:DIG_scatter}
\end{figure*}

In Section \ref{ssec:validation}, we showed that the geostatistical models constructed via the methods described in Section \ref{ssec:fitting} may be used to produce accurate predictions of the metallicities of \Hii regions.
We now use the method of universal kriging to predict the metallicities of DIG-dominated regions, and compare these predictions to the metallicities that are measured using the \Hii region calibrated metallicity diagnostics listed in \ref{ssec:Z-diagnostics}. 
In Figure \ref{fig:DIG_scatter}, the metallicities predicted by this model ($Z_{\textrm{pred}}$) are compared to the metallicities measured by applying the strong emission-line calibrations to the spectra of the DIG-dominated regions ($Z_{\textrm{meas}}$). In short, we find that generally they don't agree. For all diagnostics, values of $Z_{\textrm{pred}}$ are almost always greater than $Z_{\textrm{meas}}$. This implies that either (i) the metallicity of the DIG is intrinsically lower than that of \Hii regions, (ii) our geostatistical model is not accurate, or (iii) that traditional strong emission line diagnostics under-predict the metallicity of the DIG by a factor of $\sim 0.2$ dex. We believe that this final option is the most likely explanation, although we cannot rule out the first option. Depending on the efficiency of metal mixing throughout the ISM, \Hii regions may be more likely to be metal-rich than DIG-dominated regions, due to star-formation being more likely to trigger in higher-metallicity regions (e.g. \citealt{Wolfire+03}), and the recent release of metals from the deaths of massive stars creating local enrichment around the sites of the stars' birth \citep{Vogt+17}. 

\begin{figure*}
    \centering
    \includegraphics[width=0.8\textwidth]{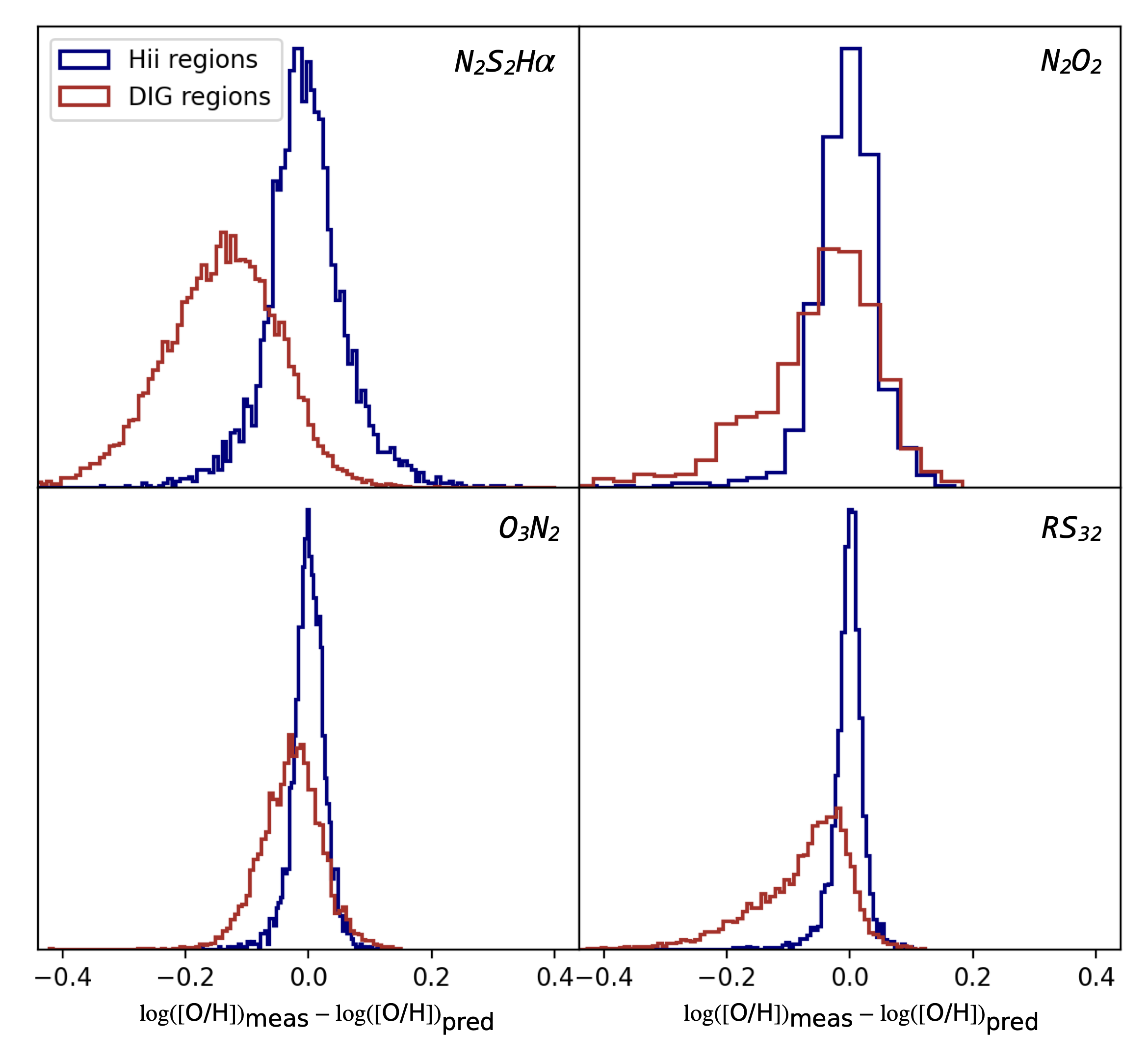}
    \caption{Histograms of the difference between the metallicities measured for each spaxel ($Z_{\textrm{meas}}$), and the metallicities predicted using universal kriging ($Z_{\textrm{pred}}$). Dark blue lines show the results of the ten fold cross-validation for \Hii regions described in Section \ref{ssec:validation}. Maroon lines show the errors in predicted vs measured metallicities for DIG regions. All diagnostics show a significant negative bias -- that is, the metallicities inferred from strong emission line diagnostics are lower than those expected from the geostatistical model.}
    \label{fig:DIG_hists}
\end{figure*}

The offset between $Z_{\textrm{meas}}$ and $Z_{\textrm{pred}}$ is revealed more clearly in Figure \ref{fig:DIG_hists}. Dark blue lines show the same histograms presented in Figure  \ref{fig:CV_hists} for the geostatistical model. Maroon lines show the offset between $Z_{\textrm{meas}}$ and $Z_{\textrm{pred}}$ for DIG-dominated regions. For all of these diagnostics, the distribution of $Z_{\textrm{meas}} - Z_{\textrm{pred}}$ for DIG regions is (i) wider than for the \Hii counterpart, (ii) negatively biased, with modes ranging from $-0.02$ (\NO) to $-0.14$ (\NSH), and (iii) skewed negatively, with longer tails in the negative direction. This is likely due to the varying amounts of DIG contamination in each spaxel. By our selection methods described in Appendix \ref{ssec:DIG-diagnostics}, all spaxels with less than $90\%$ of their $H\alpha$ emission originating from \Hii regions are classified as being ``DIG-dominated"; and metallicities determined for spaxels with higher fractions of DIG emission will show greater errors when measured with H\textsc{ii}-calibrated $Z$-diagnostics. 

Of the four diagnostics tested, we find that for \NO\ the values of $Z_{\textrm{pred}}$ for DIG-dominated spaxels are closest to the values of $Z_{\textrm{meas}}$. This agrees with the results of \citet{Zhang+17}, who state that \NO\ is largely insensitive to DIG contamination. 

\subsection{Testing DIG correction factors}
\label{ssec:vs-dig-corrections}
Using data from the MUSE Atlas of Disks (MAD), \citet{Kumari+19} computed correction factors for the \NO\ and \RS\ diagnostics that allow them to be applied to DIG-dominated regions (selected using the N2-BPT and S2-BPT diagnostics discussed in Appendix \ref{ssec:BPT-diagnostics}). Here, we detail the methodology used by \citet{Kumari+19} to construct these diagnostics, and test the metallicities measured using this method against predictions from our geostatistical model, as an independent assessment of their validity with a different sample of galaxies.

Spaxels from 24 nearby spiral galaxies observed as a part of the MAD program were classified as either \Hii or DIG/LIER regions, based on three classification schemes: the S2-BPT method, the N2-BPT method, and a surface brightness method based on a threshold flux for the H$\alpha$ brightness of \Hii regions of $\log \Sigma_{\text{H}\alpha} = 39$ erg kpc$^{-2}$ s$^{-1}$ (not considered in this work). A Voronoi-binning scheme was employed to ensure that all \Hii and DIG regions had S/N > 5 for the emission lines [O \textsc{iii}]$\lambda$5007, H$\beta$, H$\alpha$, [N \textsc{ii}]$\lambda$6584, and [S \textsc{ii}]$\lambda\lambda6717, 31$. For each Voronoi cell, metallicities were determined using the \NSH, \ON, and \RS\ diagnostics.\footnote{The \RS\ diagnostic used in \citet{Kumari+19} is not the same as the \RS\ diagnostic published in \citet{Curti+20}. Although it is computed using the same method, \citet{Kumari+19} fit only a third-order polynomial to the relationship between $Z$ and \RS, whereas \citet{Curti+20} fit a fifth-order polynomial. Furthermore, in \citet{Kumari+19}, the line ratio $\log( $[O \textsc{iii}]$\lambda5007/$H$\beta + $[S \textsc{ii}]$\lambda\lambda6717,6731$/H$\alpha )$ is referred to as O$_3$S$_2$ rather than \RS. For consistency, we will refer to this diagnostic as \RS\ throughout. In this and only this Section, we will use the third-order \RS\ calibration published in \citet{Kumari+19} and not the fifth-order calibration of \citet{Curti+20}, such that our results may be compared directly to those of \citet{Kumari+19}.} Then, pairs of \Hii and DIG/LIER regions with angular separations greater than the $1$ arcsecond, and physical separations less than $500$ pc were selected. The motivation for the first cut is that the seeing of MUSE for these observations ranges from $0.5''-1.2''$; this ensures that the correlations in metallicities between these nearby regions is not due to beam smearing. The second cut is motivated by the results of \citet{Berg+13}, who found that \Hii regions separated by less than $500$ pc had metallicity differences lower than $0.1$ dex.

\begin{figure*}
    \centering
    \includegraphics[width=0.99\textwidth]{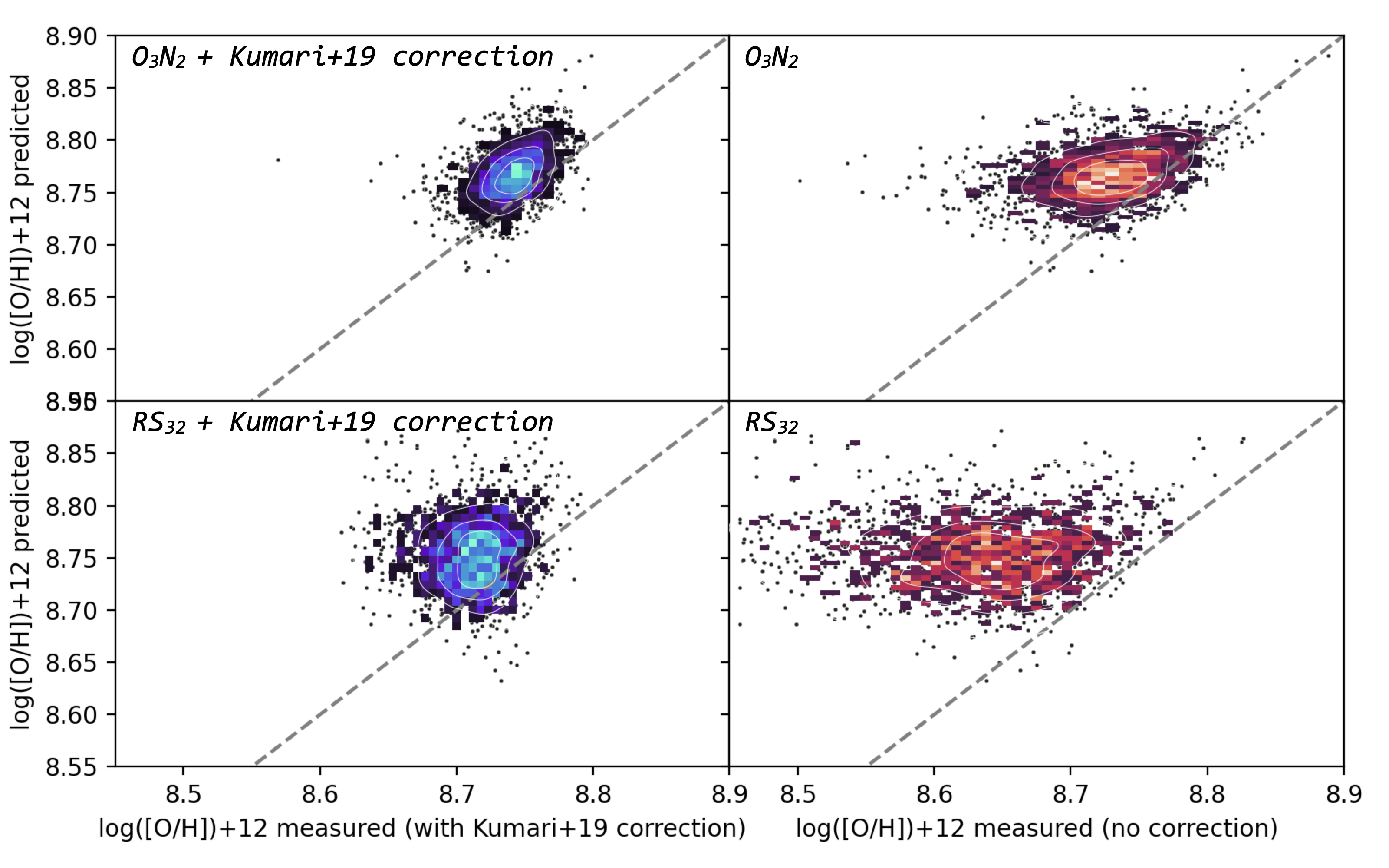}
    \caption{\emph{Left:} Scatter plot showing the difference between the metallicities predicted for DIG regions, using the geostatistical model calibrated on \Hii regions selected using the N2-BPT method, and the metallicities measured for the DIG regions using the DIG-corrected empirical diagnostics of \citet{Kumari+19}. \emph{Right:} The same, but without applying the DIG corrections to the empirical metallicity diagnostics.}
    \label{fig:Kumari_Correction_scatter}
\end{figure*}

\begin{figure*}
    \ContinuedFloat
    \centering
    \includegraphics[width=0.99\textwidth]{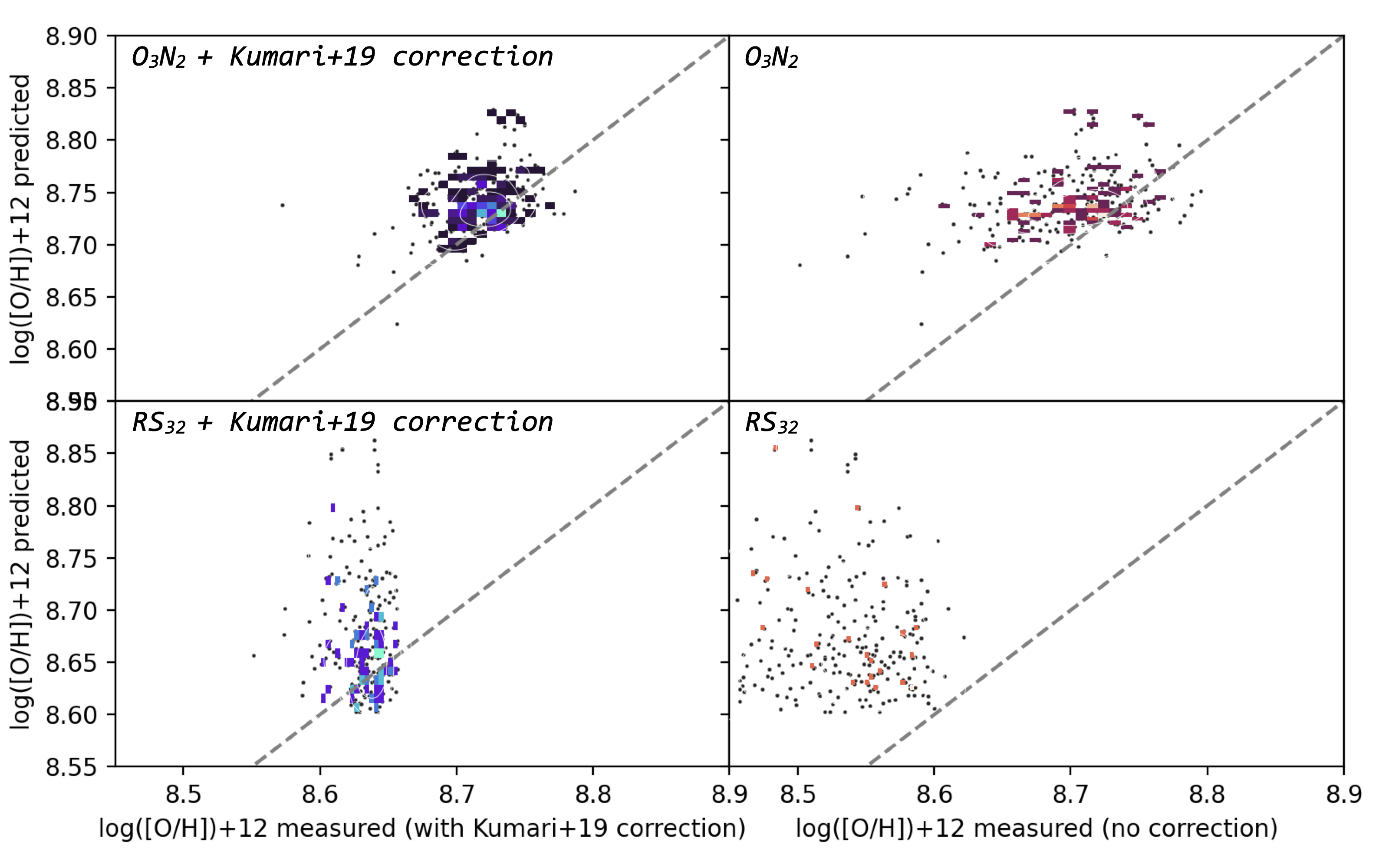}
    \caption{ -- \textit{continued}. As above, but selecting \Hii regions using the S2-BPT diagnostic.}
\end{figure*}

Using this selection of H\textsc{ii}-DIG pairs, \citet{Kumari+19} proceed using a classical (as opposed to geo)statistical analysis, assuming that the true metallicity of each DIG region in a pair should be equal to the metallicity of the \Hii region of that pair. This assumption is discussed further in Section \ref{ssec:other_studies}. Based on this assumption, $\Delta \log Z$, the difference between the metallicity measured for the \Hii region and the DIG-dominated region in each pair was calculated, and this quantity was interpreted as the error term associated with using these metallicity diagnostics to infer metallicities of DIG/LIER. Similarly to the results of our geostatistical analysis presented in Section \ref{sec:vs-dig}, they found that the differences between the metallicities measured directly for DIG regions and those inferred from the metallicities of nearby \Hii regions showed a large amount of scatter, and significant bias, especially when the \RS\ and \NSH\ diagnostics were used. Furthermore, $\Delta \log Z$ was found to correlate significantly with the line ratio O3 ($=[$O \textsc{iii}$] \lambda5007/$H$\beta)$ for \ON\ and \RS.
Using this information, \citet{Kumari+19} fit first- and second-order polynomials to the relationship between $\Delta \log Z$ and O3, using both the S2-BPT and N2-BPT DIG diagnostics, to compute correction factors for \ON\ and \RS\ that can be used to produce more accurate measurements of the metallicity of DIG.

In this Section, we use our geostatistical model, calibrated on \Hii metallicities using the \ON\ and \RS\ diagnostics, to predict the metallicity of DIG regions selected using the S2-BPT and N2-BPT methods. We compare these predicted metallicities to two measured metallicities: firstly, the metallicity of these DIG regions measured directly using the \Hii region calibrated \ON\ and \RS\ diagnostics published in \citet{Curti+17} and \citet{Kumari+19}; and secondly, adding in the O3-dependant DIG correction terms computed by \citet{Kumari+19} using the method outlined above. 

Figure \ref{fig:Kumari_Correction_scatter} shows the results of this analysis for the galaxy NGC 5236 as a scatter plot. Independently of whether the N2-BPT or S2-BPT diagnostic was used to separate \Hii regions from DIG, the metallicities of DIG regions determined using the O3-corrected empirical metallicity diagnostics were found to be closer to $Z_{\textrm{pred}}$ than those computed using the uncorrected diagnostic.\footnote{We note that less data points classified as DIG dominated when the S2-BPT diagnostic is used compared to when the N2-BPT diagnostic is used -- this is discussed further in Appendix \ref{sec:diagnostics}.} Under the assumptions that our geostatistical models are accurate (which is reasonable; see Section \ref{ssec:validation}) and the spatial structure of metallicities throughout the DIG should be the same as the spatial structure of metallicities within \Hii regions (which is reasonable, as both are components of the warm ISM, and are mixed together by the same forces), this implies that the correction factors of \citet{Kumari+19} are successful at computing more accurate metallicities for regions of galaxies for which DIG contamination is appreciable. Interestingly, the O3-corrections also reduce the $Z_{\textrm{meas}}$ produced, especially for \RS. This may indicate that these correction terms are anticorrelated with the true metallicity of the DIG regions; when added, they bring the inferred metallicities closer to the mean. More sophisticated correction factors may need to be constructed to avoid this effect.

\begin{figure*}
    \centering
    \includegraphics[width=0.8\textwidth]{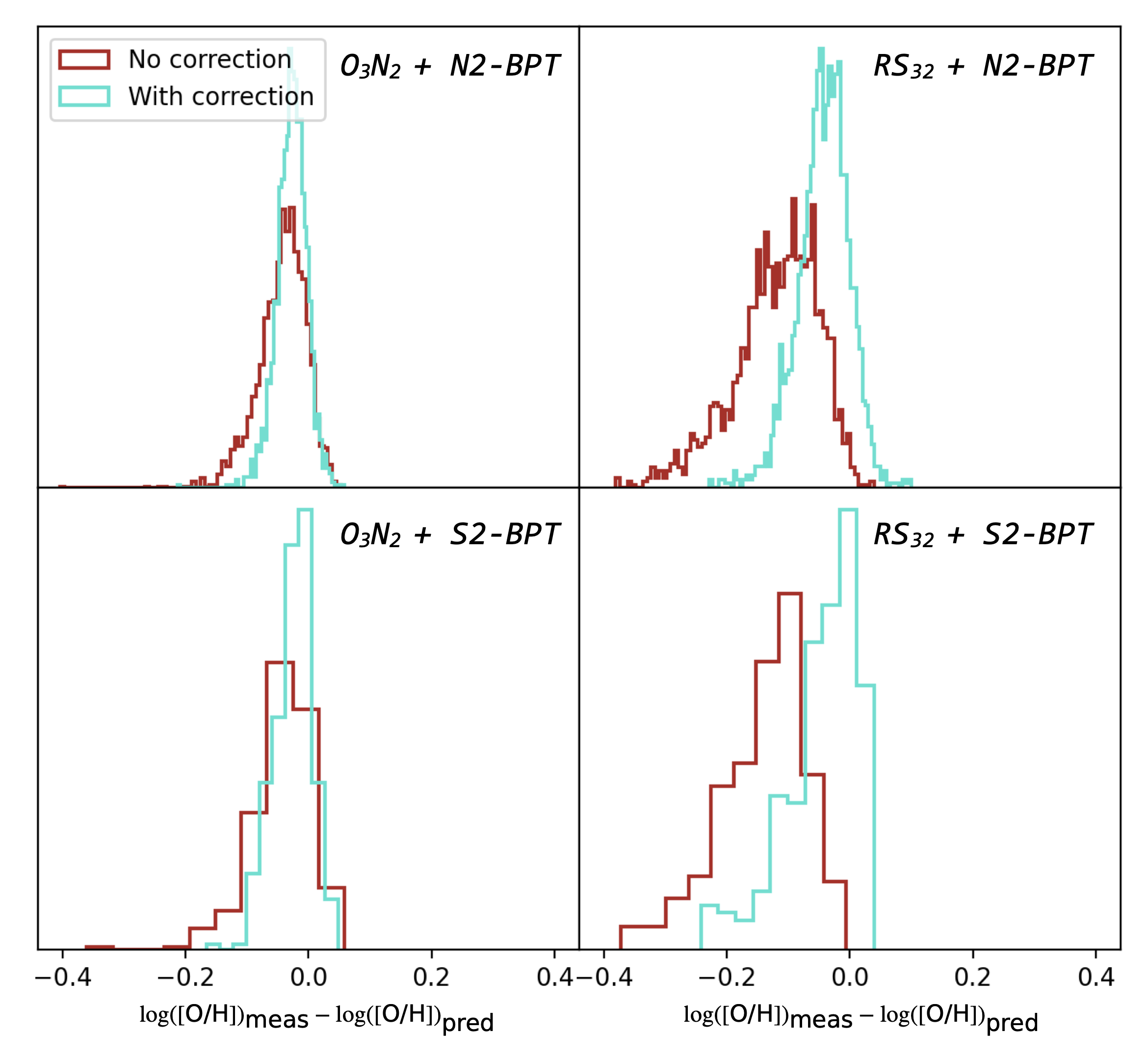}
    \caption{Histograms showing the difference between the metallicities predicted for DIG-dominated regions in NGC 5236 using a geostatistical model and those measured by the empirical metallicity diagnostics published in \citet{Kumari+19} (i) without DIG correction terms (maroon) and (ii) including the DIG correction terms (cyan). For this galaxy, the O3-dependant correction terms decrease both the bias and the scatter in $Z_{\textrm{meas}}$ -- however, this is not always the case (see Table \ref{tab:bias} and discussion in text).}
    \label{fig:Kumari_Correction_hist}
\end{figure*}

We show this same data as histograms in Figure \ref{fig:Kumari_Correction_hist}. Maroon lines are analogous to those shown in Figure \ref{fig:DIG_hists} for the N2-BPT and S2-BPT diagnostics; cyan lines show the reduction in scatter and bias gained when the O3-corrections are added. We replicate the result of \citet{Kumari+19}, in that these correction factors successfully reduce both the bias and scatter in  $Z_{\textrm{meas}} - Z_{\textrm{pred}}$. However, some bias still remains. For NGC 5236, the mean value of $Z_{\textrm{meas}} - Z_{\textrm{pred}}$ for the \ON\ diagnostic is $-0.04$ before correction, decreasing to $-0.03$ after the O3 correction factor is added (this result does not depend on whether the N2- or S2-BPT diagnostic is used). For \RS, the improvement is greater, with the mean bias in $Z_{\textrm{meas}}$ decreasing from $-0.12$ to $-0.04$ when the N2-BPT diagnostic is used (likewise, from $-0.14$  to $-0.05$ with the S2-BPT diagnostic).

\begin{table*}
    \centering
    \begin{tabular}{cr|rrrr|rrrr}
    \hline
        Galaxy & \multicolumn{1}{c}{Type} & \multicolumn{4}{c}{N2-BPT} & \multicolumn{4}{c}{S2-BPT} \\ 
        ~ & ~ & $\mu$-\ON	& $\sigma$-\ON	& $\mu$-\RS	& $\sigma$-\RS	& $\mu$-\ON	& $\sigma$-\ON	& $\mu$-\RS	& $\sigma$-\RS \\ \hline
        NGC 1068 & Before correction & -0.08 & 0.05 & -0.17 & 0.08 & -0.04 & 0.04 & -0.08 & 0.04 \\ 
        ~ & After correction & -0.02 & 0.02 & -0.10 & 0.04 & 0.04 & 0.03 & 0.07 & 0.05 \\ 
        NGC 1365 & Before correction & -0.02 & 0.05 & -0.10 & 0.07 & -0.03 & 0.05 & -0.07 & 0.04 \\ 
        ~ & After correction & 0.02 & 0.04 & -0.01 & 0.07 & 0.03 & 0.04 & 0.05 & 0.05 \\ 
        NGC 1566 & Before correction & -0.01 & 0.07 & -0.11 & 0.07 & -0.02 & 0.05 & -0.11 & 0.06 \\ 
        ~ & After correction & 0.02 & 0.06 & -0.01 & 0.06 & 0.01 & 0.05 & -0.01 & 0.06 \\ \
        NGC 2835 & Before correction & -0.01 & 0.08 & -0.07 & 0.05 & -0.03 & 0.06 & -0.06 & 0.04 \\ 
        ~ & After correction & 0.06 & 0.06 & -0.06 & 0.04 & 0.01 & 0.05 & 0.06 & 0.05 \\ 
        NGC 2997 & Before correction & -0.02 & 0.06 & -0.09 & 0.06 & -0.04 & 0.05 & -0.12 & 0.06 \\ 
        ~ & After correction & 0.02 & 0.05 & 0.01 & 0.06 & -0.01 & 0.04 & -0.02 & 0.07 \\ 
        NGC 5068 & Before correction & -0.01 & 0.06 & -0.08 & 0.05 & -0.04 & 0.06 & -0.08 & 0.04 \\ 
        ~ & After correction & 0.05 & 0.04 & 0.06 & 0.04 & 0.00 & 0.04 & 0.03 & 0.05 \\ 
        NGC 5236 & Before correction & -0.04 & 0.04 & -0.12 & 0.07 & -0.04 & 0.05 & -0.14 & 0.07 \\ 
        ~ & After correction & -0.03 & 0.02 & -0.04 & 0.04 & -0.03 & 0.03 & -0.05 & 0.06 \\ 
        NGC 7793 & Before correction & -0.01 & 0.07 & -0.07 & 0.05 & -0.04 & 0.05 & -0.08 & 0.04 \\ 
        ~ & After correction & 0.04 & 0.06 & 0.05 & 0.04 & -0.01 & 0.04 & 0.01 & 0.05 \\ \hline
    \end{tabular}
    \caption{Mean offset $(\mu$) and scatter $(\sigma)$ in $Z_{\textrm{meas}} - Z_{\textrm{pred}} $ found for DIG regions for each galaxy, when using the corrected vs uncorrected \ON\ and \RS\ metallicity diagnostics.}
    \label{tab:bias}
\end{table*}

Unfortunately, this trend of decreasing total bias was not seen for all galaxies in our sample. In Table \ref{tab:bias}, we document the mean values and scatter in of $Z_{\textrm{meas}} - Z_{\textrm{pred}}$ for DIG regions in each galaxy, selected using both the N2- and S2-BPT diagnostics, comparing the amount of bias before and after the O3 corrections are applied. We find that generally the bias is lower when the DIG-correction factors are used. However, several cases can be seen where the O3-based correction factor introduces a larger bias than the uncorrected diagnostic. Furthermore, the scatter in $Z_{\textrm{meas}} - Z_{\textrm{pred}}$ often is just as wide when the corrected diagnostics are used as when the uncorrected diagnostics are used. We conclude that while the O3-based correction factors of \citet{Kumari+19} are successful in improving predictions of metallicities for regions with significant DIG contamination, these correction factors are not universal. Indeed, different galaxies may host different sources of DIG ionisation, including evolved stars, shocks, and leaking \Hii regions. If this were the case, then finding a single correction factor to negate the effects of DIG contamination may not be possible, due to the different sources of DIG ionisation in each galaxy. We discuss this point further in Section \ref{sec:discussion}. 

\begin{figure*}
    \centering
    \includegraphics[width=0.99\textwidth]{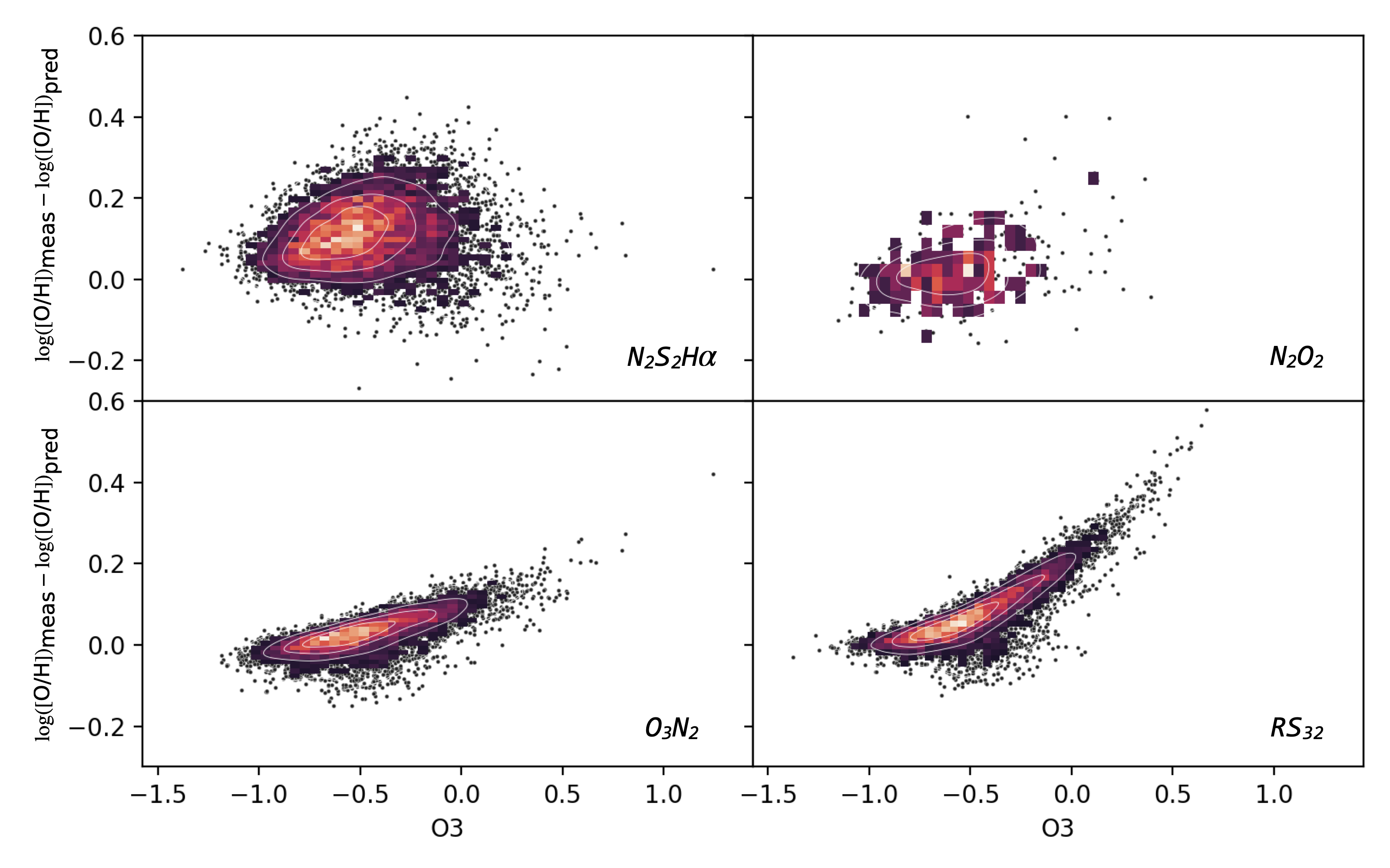}
    \caption{The deviance between metallicities measured directly using strong-emission line diagnostics for DIG-dominated regions, and those predicted via universal kriging, versus the line ratio O3$=[$O \textsc{iii}$\lambda5007]/$H$\beta$, for NGC5236, selecting \Hii and DIG-dominated regions using the S2-based method described in \citet{Kaplan+16}. For the two empirically calibrated diagnostics \ON\ and \RS, a strong correlation is seen between the error in the measured metallicities and the O3 line, in excellent agreement with the findings of \citet{Kumari+19}. For the two theoretically calibrated diagnostics, no strong correlation is observed.} 
    \label{fig:Error_vs_O3}
\end{figure*}

Finally, we demonstrate how our method of geostatistical analysis may be used to construct O3-dependant correction factors, similarly to the methodology of \citet{Kumari+19}.
In Figure \ref{fig:Error_vs_O3}, we
plot $Z_{\textrm{meas}} - Z_{\textrm{pred}}$ (using the \citealt{Curti+20} calibration for \RS) for DIG-dominated spaxels of NGC 5236, selected using the S2 method of \citet{Kaplan+16}, against the value of the line ratio O3 for our four metallicity diagnostics. We see that for the two theoretically calibrated diagnostics, there is no significant correlation. However, for both the \ON\ and \RS\ calibrations, the offset is significantly correlated with O3. The trend of $Z_{\textrm{meas}} - Z_{\textrm{pred}}$ with O3 is well-fit by a linear relation for the \ON diagnostic, and a higher-order polynomial for \RS, in qualitative agreement with the results of \citet{Kumari+19} using other DIG diagnostics. This analysis demonstrates how the results of universal kriging may be used to construct new correction factors for empirically-calibrated metallicity diagnostics to be used on DIG-dominated regions. We postpone the construction of such correction factors until the full release of data from the TYPHOON survey (Seibert et al. in prep.). 

\section{Discussion} \label{sec:discussion}

Thanks to the efforts of large population IFS galaxy surveys such as CALIFA \citep{CALIFA}, MaNGA \citep{MANGA}, and SAMI \citep{SAMI}, a clear unified picture of star-forming galaxies in the local universe has emerged. Global relationships between galaxy properties, such as the mass-metallicity relationship (MZR, e.g. \citealt{Tremonti+04}), the star-forming main sequence \citep{Renzini+Peng15}, and the Kennicutt-Schmidt relation \citep{Schmidt59, Kennicutt98} have been found to be large-scale consequences of local scaling laws \citep[e.g.][]{Moran+12, Rosales-Ortega+12, Sanchez+13, Wuyts+13, Bigiel+08}. Furthermore, observational evidence suggests that galaxies usually evolve inside-out, with inner regions showing higher star formation rates, metallicities, gas densities, and stellar ages \citep{Sanchez20}.

This picture holds over a large range of masses and morphologies. However, in this work, we show that the simple linear metallicity gradient is far from the whole picture. The failure of simple azimuthally symmetric models to capture the complex structures of the ISM of galaxies has been noted explicitly by many studies, especially at high redshift, and when dealing with high-resolution data. In a study of strongly lensed galaxies at cosmic noon ($1.2 < z < 2.4$), \citet{Curti+20b} find that the process of fitting a metallicity gradient can often obscure kpc-scale inhomogeneities that are common in starbursting systems. Similarly, \citet{Florian+21} suggest that star-formation in strongly-lensed galaxies at $z \sim 1.4$ is better characterised by its `clumpiness' rather than a large-scale gradient. 

Our geostatistical approach is useful for describing both the large scale radial trend and the small-scale clumpy structure of the ISM in a parsimonious way, using only four parameters. Furthermore, this approach is based on over 60 years of literature, with applications ranging from agriculture to epidemiology. 

For all of the galaxies in our extended sample of eight nearby galaxies observed using the TYPHOON/PrISM survey, we find that galaxies show stochastic variations over characteristic spatial scales of $\sim 1$ kpc. This is seen regardless of the metallicity and DIG diagnostics used to construct \Hii metallicity maps. In Section \ref{sec:results}, we show that even a simple exponential model of the stochastic small-scale metallicity structure is able to account for a large amount of the variance found around the large-scale metallicity trend. Furthermore, in Section \ref{ssec:validation}, we show that by accounting for these small-scale fluctuations, the accuracy of the prediction of the metallicity at all regions of a galaxy can be vastly improved, especially when the metallicity gradient of the galaxy is close to zero. This makes this technique particularly appealing for the analysis of high-redshift galaxies, where most galaxies do not exhibit large-scale metallicity gradients \citep{CLEAR}. Results of our universal kriging procedure will be presented for the seven currently unpublished galaxies listed in Table \ref{tab:obs_table} in a future paper (Metha et al. in prep.).

This technique has potential applications in many areas of extragalactic astronomy where accurate metallicity predictions are useful. In particular, this technique may be valuable in determining the value of the Hubble constant, $H_0$, using observations of the local universe. Type Ia supernovae (SNe Ia) are the prototypical standard candles, with a narrow scatter in their intrinsic luminosity \citep{Folatelli+10}. These bright transient events can be used as a cosmological distance ladder when calibrated using other standard candles, such as RR Lyrae variables and Cepheids, which have well-defined period luminosity (PL) relations. However, these PL relations are known to depend on metallicity \citep{Nemec+94, Groenewegen+04}. To calibrate these relations, accurate predictions of the metallicity of the interstellar gas at the locations of these stars are needed. Metallicity gradients alone have a large prediction variance at all radii, which can contribute significantly to the error budget of the cosmological distance ladder \citep{Beaton+16}. 
In Figure \ref{fig:krig}, we show that this prediction variance can be decreased when universal kriging is used. These more accurate predictions would lead to a more accurate calibration of the zero point of the SNe Ia luminosity function, leading to tighter constraints on the value of $H_0$ in the local universe. 

In order to compute high-resolution molecular hydrogen column density maps for nearby galaxies, the PHANGS-ALMA survey must estimate the metallicity dependant CO-to-H$_2$ conversion factor, $\alpha_{CO}$. In \citet{Sun+20}, this is done by assuming a linear gradient model, with $Z_c$ estimated from a galaxy's stellar mass using the MZR, and a constant assumed metallicity gradient of 0.1 dex $R_e^{-1}$ for all galaxies. Therefore, estimates of the molecular gas density distribution could be improved upon by using these methods, together with PHANGS-MUSE data. We emphasise that our universal kriging method could provide metallicities at all locations of the galaxies, even at regions of DIG contamination.

As telescope instrumentation continues to improve, more and more IFS surveys are being conducted which allow the ISM of nearby galaxies to be resolved to scales of $100$pc or less, below the size of an individual \Hii region \citep{Mannucci+21}. Recent and upcoming surveys with resolutions similar to TYPHOON/PrISM include PHANGS \citep{Emsellem+21}, MAD \citep{Erroz-Ferrer+19}, the SDSS-V LVM surveys of the LMC and SMC \citep{Kollmeier+17}, TIMER \citep{Gadotti+19}, SIGNALS \citep{SIGNALS}, GASP \citep{Poggianti+17}, and AMUSING \citep{Galbany+16}. The geostatistical techniques described in this work are naturally applicable to this new generation of IFS surveys. Comparisons between the small-scale metallicity structure of the ISM of these galaxies and the output of dedicated zoom-in comparison simulations \citep{Jeffreson+20, Utreras+20} will allow us to constrain models of metal mixing through stellar feedback. Furthermore, these techniques may be used in conjunction with data from the TIMER survey to reveal the effects of spiral bars and nuclear structures on a galaxy's internal metallicity distribution, or the GASP survey to better understand the mixing effects in interacting galaxies.

In Figure \ref{fig:Error_vs_O3}, we show that the line ratio O3 correlates with the offset between $Z_{\textrm{meas}}$ and $Z_{\textrm{pred}}$ for two empirically calibrated metallicity diagnostics, confirming the result of \citet{Kumari+19}. Current state-of-the-art Bayesian metallicity fitting programs such as IZI \citep{Blanc+15} are not able to correct for DIG contamination in regions \citep{Zhang+17}. Our Figure \ref{fig:Error_vs_O3} implies that the O3 line ratio may be useful in correcting for DIG contamination in order to produce a robust metallicity fit.

\subsection{Comparisons to other works}
\label{ssec:other_studies}

This study is not the first that attempts to predict the metallicity throughout the ISM using data from \Hii regions. In this Section, we briefly review some other key studies, and highlight how our analysis differs from the current state of the art.

Using data from 365 galaxies from the MANGA survey, \citet{Zhang+17} derive the bias occurring from DIG contamination associated with different strong emission line based metallicity diagnostics. 
Assuming that the proportion of DIG contamination within a spaxel is proportional to the surface brightness of H$\alpha$, a linear relation was fit between the metallicity derived using several strong-line diagnostics, and the amount of DIG contamination.
The effect of metallicity gradients was avoided by only selecting spaxels within $0.4-0.6R_e$ within each galaxy.
No local metallicity variations or metallicity measurement errors were accounted for. In the language of our geostatistical model, this would be equivalent to setting $\eta(\vec{x}) = 0$ and $\epsilon(\vec{x}) = 0$. 
Out of seven diagnostics tested, \citet{Zhang+17} concluded that the two least sensitive to contamination from DIG were the \NSH\ diagnostic with the calibration of \citet{Dopita+16}, and the \ON\ diagnostic with the calibration of \citet{Pettini+Pagel04}.

This method, while statistically robust, does not account for any local variations in metallicity. For this reason, a substantial fraction of the data need to be discarded in order to make this technique work. In contrast, by using geostatistical modelling, all the data in the galaxy could be used, leading to a larger sample of spaxels, which could strengthen the conclusions of such a study.

Similarly, the approach of \citet{Kumari+19} can be casted as a geostatistical analysis, but with a very specific form for the correlation function. Specifically, the correlation in metallicity measurements at two different data points is treated as a top hat function: $\text{Corr}(Z(\vec{x}),Z(\vec{y})) = 1 $ if $| \vec{x} - \vec{y} | < 0.5$ kpc, and $0$ otherwise. This correlation structure would induce a shape for the theoretical semivariogram of the following form for all galaxies:

\begin{equation}
    \gamma(h) = \begin{cases}
      0, & \text{if}\ h<0.5 \\
      \sigma^2, & \text{otherwise}
    \end{cases}
    \label{eq:Kumari_svg}
\end{equation}

Our Figure \ref{fig:semivariograms} shows that such a model is not a good fit for local star forming galaxies. Furthermore, the method of \citet{Kumari+19} does not account for any large-scale metallicity trends, such as metallicity gradients. Finally, such a step-based correlation function does not lead to a positive definite covariance function.\footnote{This can be seen from Bochner's theorem: a Fourier transformation of a top-hat function yields a sinc function, which is not positive for all real numbers.} Therefore, the model of \citet{Kumari+19} for describing the small-scale homogeneity of the ISM is not statistically valid, and its application may thus lead to hard to quantify biases in the inference drawn from it.

Using a hierarchical model optimised with the Integrated Nested Laplace Approximations method (INLA, \citealt{Rue+17}), \citet{Gonzalez-Gaitan+19} fit a model of the spatial variation of the age and metallicity of stellar populations in 721 galaxies from the CALIFA and PISCO \citep{Galbany+18} surveys. The model that they fit is similar to our own in many respects: in both cases, the mean trend $\mu$ is modelled as being linearly dependant on the radial distance from the galaxy centre; and in both cases, the observed metallicity at each location is assumed to be drawn from a latent Gaussian process with some known observation error. Our analysis differs from theirs in two key respects: firstly, in that work the covariance kernel of the small-scale metallicity variation $\eta$ is a Mátern function, rather than an exponential function, allowing one extra free parameter to be fit.\footnote{The Mátern function is a stationary positive definite correlation function containing two parameters: the order $\nu$, which sets the shape of the function, and the range, $\phi$. The exponential function is a special case of the Mátern function where $\nu = 0.5$.} Secondly, the fitting method that they use is Bayesian, whereas ours uses a maximum likelihood approach. Using a Bayesian method may make the results dependant on the model priors, and be computationally costly to compute, even though the speed of computations is significantly increased by using the INLA approximate Bayesian computing method. 

Similarly, \citet{Clark+19} also use Bayesian methods to fit a Gaussian process model of the metallicity variation within galaxies, using a Mátern function of order 1.5 to capture deviations from a mean radial trend. Using a cross-validation analysis (their Appendix C), they found that this method produced accurate predictions of the metallicities of \Hii regions, with $80.5\%$ ($85.4\%$) of predicted \Hii region metallicities agreeing with their measured metallicity values for M74 (M83) -- larger than the $68.3\%$ proportion expected for a Gaussian process. 
This result may be caused by an overestimation of the uncertainty of the difference between the predicted metallicity and the actual metallicity: in their equation C1, the variance of this difference is derived under the assumption that the predicted metallicity and the actual metallicity are independent of each other at each testing point. However, since the predicted metallicity is constructed using the metallicities observed, and these observed metallicity are spatially correlated with the true metallicity at each testing point, the predicted metallicity and the actual metallicity are likely to be correlated.
Accounting for this positive correlation between predictions and the true metallicity would likely reduce or remove this tension.

The methodology of \citet{Williams+22} follows \citet{Clark+19} very closely, fitting the same model using the same software (\texttt{GaussianProcessRegressor} from the \textsc{python} machine learning package \texttt{scikit-learn}), and focusing on analysing the results of their best-fit models in order to understand the spatial scales over which galaxies may be considered well-mixed. The length scales of best-fit for their Mátern kernels range from $1.6-30$ kpc for the 12 galaxies with significant metallicity variations beyond a radial trend (see left panel of Fig.~9 in \citealt{Williams+22}), with a median length scale of $\sim 16$ kpc, comparable to the size of the galaxy. This length scale is much larger than the $\sim 0.1-0.4$ kpc mixing scales we found with our exponential kernel. \citet{Williams+22} also computed the $50\%$ correlation scale for these PHANGS galaxies, finding mixing lengths of $0.2-1.1$ kpc (see right panel of Fig.~9 in their work), that are in agreement with our findings. 
Therefore, we conclude that it is not the data, but the modelling methodology of \citet{Williams+22} that leads to the observation of non-radial metallicity variations on a different physical scale to the model presented here. In our analysis, we separate the signal $Z(x)$ into a large-scale global trend ($\mu(x)$, e.g. a metallicity gradient) and small-scale fluctuations ($\eta(x)$; characterised through the semivariogram analysis), whereas the models of \citet{Williams+22} capture galaxy-scale trends that are nonetheless significantly different from a simple metallicity gradient-based approach. We emphasise that the small-scale fluctuations we uncover in our work cannot be consistent with measurement error, as they are spatially correlated on scales of $\sim 1$ kpc, whereas measurement error is expected to have no spatial correlation.

In our work, we model the small-scale stochastic structure of the ISM as a stationary, isotropic random field. Other models \citep[e.g.][]{Sale+Magorrian14, Lee+Gammie21} relax this assumption, employing non-stationary models that allow parameters such as the correlation length scale to vary throughout a galaxy. This procedure introduces additional complexity that makes inference more challenging. We find that a stationary process with a constant mixing scale length, while perhaps less realistic, is sufficient to reproduce accurately the metallicity substructures observed in galaxies, for the purpose of making reliable predictions (see Section \ref{ssec:validation}). 

All of these examples show how our novel geostatistical techniques provide a complementary approach to existing literature approaches to spatially resolved studies of metal distribution in galaxies, and offer distinct advantages.  

\subsection{Caveats}

No model exists without assumptions. In this Section, we detail the assumptions that we make in constructing our geostatistical model, and our justifications for making them.

Crucially, this analysis assumes that the metallicity of DIG-dominated regions is governed by the same physical processes that determine the metallicity of \Hii regions. This assumption allows us to use a geostatistical metallicity model trained on \Hii region data to estimate the metallicities of DIG-dominated regions. If there was a reason to believe that the metallicity of DIG regions was expected to be intrinsically lower or higher than the metallicity of \Hii regions, this bias would need to be accounted for before any metallicities could be predicted. Based on current evidence, we see no physical reason as to why this would be the case. Both the DIG and \Hii regions exist in the warm phase of the ISM. Mixing effects due to interstellar turbulence, bar driven motion, and other phenomena that act to redistribute metals through the ISM should not be significantly affected by the ionisation phase of the medium. Furthermore, the model we fit for $\eta$, the small scale metallicity structure of the ISM, is so simple that it will likely not be affected by accounting for these effects, if they exist.

This analysis also ignores the multiphase structure of the ISM and assumes each phase has a similar mixing scale. This assumption may not be accurate (see e.g. \citealt{deCia+21}). Gas particles in the hot phase of the ISM have higher velocities than warm and cold ISM constituents, which may lead to shorter timescales and spatial scales for mixing. For a first approximation, we assume that an exponential kernel with a single scale length is sufficient to capture the overall behaviour of the ISM. We will explore this assumption by analysing the spatial distribution of other ISM components in future papers.

Thirdly, this analysis does not account for aperture effects. \citet{Mannucci+21} posit that this might be a problem. As the size of our spaxels is often below the size of a \Hii region, some of our spaxels may only include the inner regions of \Hii regions, where higher ionisation species dominate, and lower ionisation species are under-represented. A tradeoff exists between integrating fluxes between \Hii spaxels to get more accurate metallicities, and losing resolution because of binning. \Hii region detection software such as \texttt{HiiPhot} (\citealt{Thilker2000}, used by \citealt{Williams+22}) or \texttt{HiiExplorer} \citep{Sanchez+12} may be used to construct integrated spectra of \Hii regions; however, these software packages rely on multiple parameters that are tuned by the user, and may convolve multiple nearby poorly-resolved \Hii regions (e.g. \citealt{Grasha+22}). Since we are trying to understand how the metallicity changes between \Hii regions that are separated by a small distance, taking this step may limit the ability to measure the shape of the semivariogram at low separation, impacting our abilities to construct geostatistical models. While aperture effects may lead to inaccuracies in metallicities computed from emission-line ratios, we do not believe these effects will be the same for all four metallicity diagnostics tested in this work. Therefore, any conclusions that hold across all diagnostics may still be trusted.
Exploring how our results change when integrated \Hii region properties are used may be the focus of future work. 

Finally, this model assumes that the stochastic metallicity fluctuations $\eta(\vec{x})$ are Gaussian, with an exponential correlation function. Of all the assumptions that we make in this work, we regard this assumed correlation structure as the main area for future improvement. From the semivariograms presented in Figure \ref{fig:semivariograms}, we can see that this model does a good job of describing the small-scale metallicity fluctuations; but it is not perfect. Below 1 kpc, semivariograms were seen to exhibit power-law like behaviour that is not captured by a simple exponential model, but is predicted from general models of turbulence.
High-resolution datacubes such as the ones produced by the TYPHOON/PrISM survey allow testing and verification of detailed models of chemical mixing in the ISM. The question of how the small-scale structure of metals in the ISM of local star-forming galaxies should best be modelled remains open and is an active area of investigation.

\section{Summary and conclusions} \label{sec:conclusions}

In \citetalias{Metha+21}, the potential of geostatistical analysis was introduced as a tool to (1) understand the small-scale structures of galaxies, (2) compare theoretical models to high-resolution IFU data, (3) constrain models of stellar feedback and metal mixing, and (4) move beyond the metallicity gradient to understand the 2D metallicity distribution of galaxies. Here, we extend upon the initial work, introducing analytical techniques for fitting a geostatistical metallicity model to galaxy data, and using these models to predict the metallicity at unmeasured points using universal kriging. We summarise our key findings below:

\begin{enumerate}
    \item For NGC 5236, significant correlations are seen in the metallicity of spaxels separated by less than $0.4-1.2$ kpc, with the scale of these correlations dependant on the metallicity diagnostic used. Small-scale metallicity structures are shown to cause deviations in the metallicities of spaxels from a radial metallicity trend with amplitude $\sim 1$  order-of-magnitude larger than what is expected from measurement errors alone. This small-scale metallicity structure appears to be fairly well-described by an exponential covariance structure with a scale parameter of $\phi=133-400$ pc and a variance of $\sigma^2 = 0.0013-0.01$; however, we believe a more physically motivated model could be constructed in the future.
    \item Universal kriging produces better predictions for metallicities at unknown data points than the commonly-used linear metallicity gradient model. Small-scale metallicity trends are important for understanding the structure of the ISM of local galaxies and thus quantifying them gives distinct advantages for testing models of turbulent metal mixing in galaxies, and determining the metallicities of specific locations within galaxies, such as the sites of transient events.
    \item The metallicity of a DIG-dominated region may be estimated using the technique of universal kriging, together with the geostatistical modelling methods presented in this work. We argue that this method is to be preferred over directly applying \Hii region calibrated metallicity diagnostics.
    \item The \NSH, \NO, \ON, and \RS\ diagnostics have limited accuracy in determining the metallicities of DIG-dominated regions. Out of the 4 metallicity diagnostics tested, \NO\ is the least sensitive to DIG contamination, in concordance with the findings of \citet{Zhang+17}.
    \item For \ON\ and \RS, the offset between the metallicities inferred from a geostatistical model via kriging and from those estimated using these \Hii-region calibrated diagnostics is tightly correlated with O3. 
    \item Comparisons to predictions from geostatistical models calibrated on \Hii region data reveal that the \ON\ and \RS\ diagnostics together with the O3-dependant DIG correction factors of \citet{Kumari+19} are found to generally exhibit less bias than their uncorrected counterparts. However, we caution that for some galaxies, these corrected metallicity diagnostics still introduce a greater bias than when no correction is used. This suggests that the dominant sources of DIG may be different for different galaxies, again highlighting the benefits of taking a more rigorous geostatistical approach to data analysis. 
\end{enumerate}

\section*{acknowledgements}

The authors thank the anonymous referee for their comments, which helped improve the structure and readability of this paper. We would like to give thanks to Prof. Barry F. Madore (Carnegie Observatories and the University of Chicago) PI of the TYPHOON Programme, to members of that team, especially Dr. Henry Poetrodjojo, Prof. Lisa Kewley, Dr. Mark Seibert and Dr. Jeff Rich for providing the data and numerous insights; to Dr Alex Cameron for his expertise on BPT diagnostics; and to Dr. Nicha Leethochawalit for their helpful comments in preparing this manuscript. 
BM acknowledges support from an Australian Government Research Training Program (RTP) Scholarship. This research is supported in part by the Australian Research Council Centre of Excellence for All Sky Astrophysics in 3 Dimensions (ASTRO 3D), through project number CE170100013.

\section*{data availability}

Any data products presented in this work, as well as the Python code used to construct them, are available from the corresponding author upon reasonable request.

\bibliographystyle{mnras}
\bibliography{biblio} 

\newpage
\appendix

\section{Comments on other TYPHOON galaxies}
\label{ap:other_galaxies}

In this paper, we largely restrict our discussion to NGC 5236, using the DIG-diagnostic procedure of \citet{Kaplan+16}. To ensure our results are valid for a wider variety of galaxies, we repeat our analysis for a selection of seven other galaxies for which high-resolution metallicity data has been obtained by the TYPHOON/PrISM survey. Properties of this extended sample of galaxies are listed in Table \ref{tab:obs_table}. 

\begin{figure*}
    \centering
    \includegraphics[width=\textwidth]{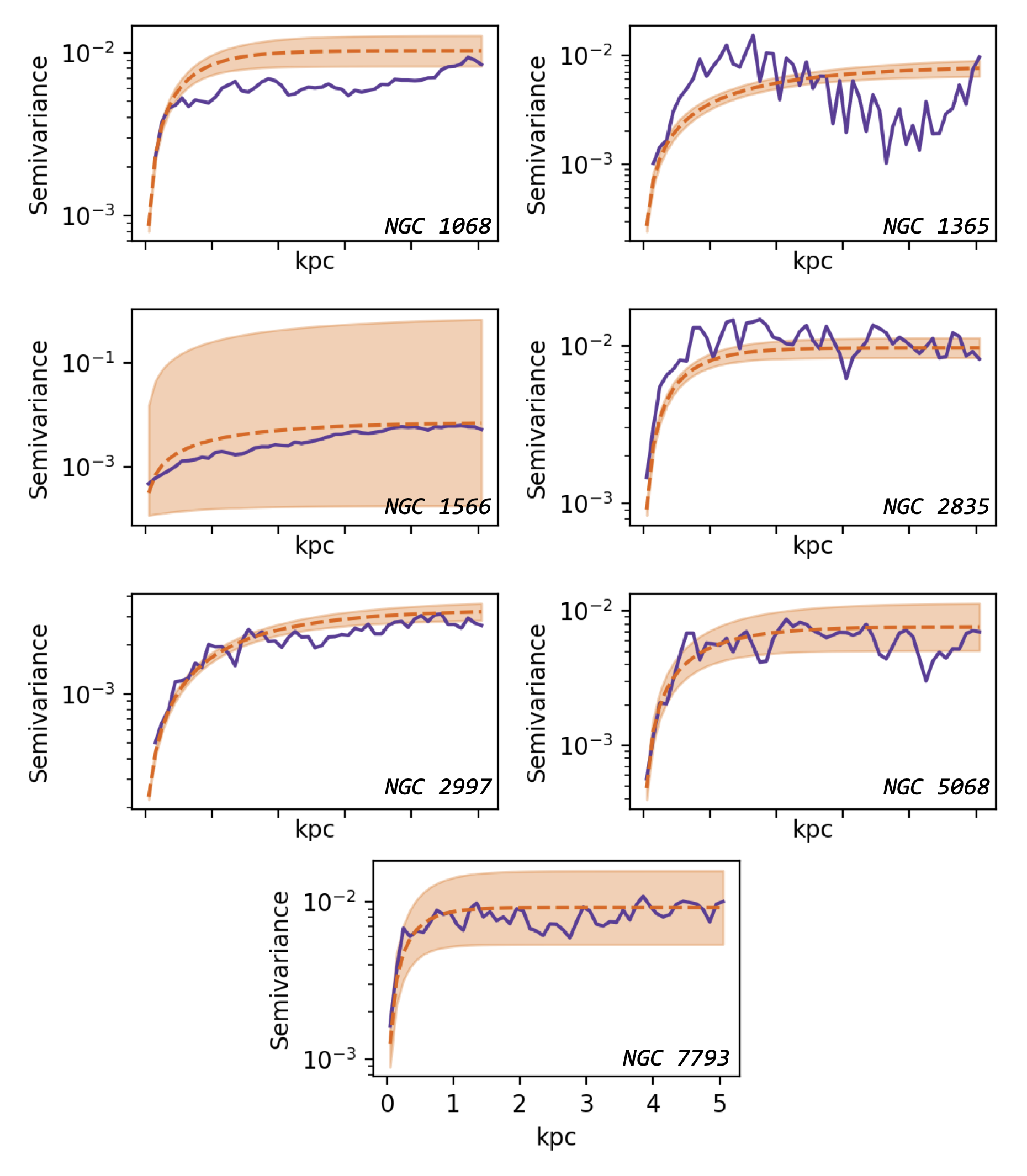}
    \caption{Semivariograms showing the small-scale metallicity structure of all other galaxies investigated in this work, using the \RS\ diagnostic. Purple solid lines show the empirical semivariogram. Orange dashed lines show the semivariogram predicted from the model of best fit. Orange shaded regions show the $1-\sigma$ error in the fitted curve (errors on the semivariance estimation from the data at each separation shown are negligible). For NGC 1365, an exponential model does not fit the data (likely a simple metallicity gradient does not subtract off all of the large-scale trends), and for NGC 1566, the uncertainties in the best-fitting model are very large. In all other cases, we see that a simple exponential model does a fair job of fitting the data, and correlations are present between data points separated by $\sim 1$ kpc.}
    \label{fig:SV_supp}
\end{figure*}

Broadly, conclusions found for NGC 5236 hold for the other seven galaxies in the sample, using any of the DIG isolation methods discussed in Appendix \ref{ssec:DIG-diagnostics} (except for the effect of applying O3-based DIG correction factors to the empirically-calibrated metallicity diagnostics of \citet{Curti+20}; see Section \ref{ssec:vs-dig-corrections}). Semivariograms constructed for all galaxies reveal significant structures in the metallicity distributions, with correlations on spatial scales of $\sim 1$ kpc. For all galaxies, the size of these correlations was much larger ($\sim 1$ order of magnitude) than the uncorrelated measurement error. In all cases, the method of universal kriging as described in the text was verified to produce more accurate predictions of the metallicities of unknown regions than a simple metallicity-gradient model using 10-fold cross-validation (Metha et al. in prep.). 
Figure~\ref{fig:SV_supp} shows the semivariograms for all galaxies in our sample for the \RS diagnostic to illustrate that in all cases a consistent correlation length is inferred. Full plots analogous to those presented in Figures~\ref{fig:semivariograms}-\ref{fig:krig} for all galaxies with all combinations of DIG/metallicity diagnostics and kriged maps showing the metallicity distribution of each galaxy will be released in a forthcoming paper (Metha et al. in prep.).

One notable exception was found when the geostatistical model-fitting method described in Section \ref{ssec:fitting} was applied to \NO\ based metallicity maps for \Hii regions selected using either the N2-BPT or S2-BPT methods for the 7 galaxies in this sample other than NGC 5236. When this was done, the amount of uncorrelated variance revealed by a semivariogram analysis was much greater than the amount predicted by our geostatistical model. We recall that our model explains all metallicity variance within a galaxy using a combination of three factors: a large-scale metallicity trend ($\mu$, modelled here as a radial metallicity gradient), small-scale metallicity fluctuations ($\eta$), and measurement errors stemming from uncertainties in the emission-line fluxes ($\epsilon$, computed using Equation \ref{eq:Z_error}). Only the measurement error term, $\epsilon$, is capable of changing the semivariance at zero separation, implying that the failure in our model to match the data must lie in an underestimation of this term. We further observe that the semivariograms computed using the \NSH\ and \ON\ diagnostics show good agreement between the modelled values of $\epsilon$ and the revealed semivariogram, implying that the problem must not lie in the error reporting associated with the [N \textsc{ii}] emission line. Therefore, we conclude that the failure of our geostatistical model to correctly model the uncorrelated variance of \NO\ is possibly due to under-reporting of the error associated with the [O \textsc{ii}] emission line. This issue will be corrected before kriged metallicity maps for the full TYPHOON sample are released to the public (Metha et al. in prep.).

This analysis highlights the power of the semivariogram analysis in being able to separate small-scale metallicity variations from measurement errors. While we focus in this study, and in \GeoGalsI, on using this analysis to understand the small-scale structure of metallicity fields after isolating and subtracting the effects of measurement error, these methods can also be used to understand errors associated with metallicity measurement without the interfering effects of true spatial variation of the data. This may be useful for validating data pipelines, and testing assumptions about the size and correlation structure of errors between spaxels in IFS data.

\section{Strong emission line diagnostics}
\label{sec:diagnostics}

Many different diagnostics exist that can be used to estimate metallicities from ratios of strong-emisssion lines in \Hii regions (e.g. \citealt{Kewley+19, MaiolinoMannucci19}). Similarly, many different diagnostics are currently being developed in order to distinguish regions of galaxies dominated by DIG emission from \Hii regions, such as using the surface brightness of H$\alpha$ emission \citep{Reynolds98, Zhang+17}, the equivalent width of the H$\alpha$ emission line \citep{Lacerda+18, ValeAsari+19}, BPT diagrams \citep{BPT, Kumari+19}, and the [S \textsc{ii}]/H$\alpha$ line ratio \citep{Reynolds85, Blanc+09, Kaplan+16,Poetrodjojo+19}. In this section, we will describe the four metallicity diagnostics and three DIG diagnostics used in this work to distinguish \Hii regions from DIG-dominated regions and identify the metallicities of \Hii regions, and how these diagnostics are calibrated.

\subsection{Metallicity diagnostics}
\label{ssec:Z-diagnostics}

In this study, we consider four different metallicity diagnostics, two of which have been calibrated theoretically using photoionisation models, and two of which are empirically calibrated against metallicities determined using the direct ($T_e$-based) method.

The primary metallicity diagnostic we consider is the \NSH\ diagnostic, devised by \citet{Dopita+16}, theoretically calibrated using the \textsc{Mappings 5.0} code \citep{Sutherland18}. The three emission lines used in this diagnostic ([N\textsc{ii}$]\lambda$6583, [S\textsc{ii}$]\lambda\lambda$6717,6731, and H$\alpha$) are all very close to each other in wavelength, reducing the sensitivity of this diagnostic to dust extinction. Furthermore, this diagnostic shows little dependence on the ionisation parameter.

As a secondary theoretically-calibrated metallicity diagnostic, we use the \NO\ diagnostic, based on the line ratio $\text{\NO} = \log ( $[N \textsc{ii}]$\lambda6583/$[O \textsc{ii}]$\lambda\lambda3726,29)$. \citet{Zhang+17} found that metallicities calculated using this diagnostic showed very little dependence on DIG contamination. To match this study, we use the theoretical calibration of \citet{Dopita+13}, which was computed using the astrophysical plasma modelling code \textsc{Mappings IV}.
An advantage of using this line ratio is that it is insensitive to variations in the ionisation parameter \citep{Dopita+00}. One disadvantage of this diagnostic for the specific dataset we used is that the du Pont Telescope at the Las Campanas Observatory is not very sensitive at the wavelength of [O \textsc{ii}]. For this reason, errors associated with this metallicity diagnostic are much larger than those found with any of the other metallicity diagnostics tested.

Additionally, we consider two metallicity diagnostics with empirical calibrations; one based on the \ON\ line ratio ($=\log( $[O \textsc{iii}]$\lambda5007/$H$\alpha ) - \log( $[N \textsc{ii}]$\lambda6583/$H$\beta )$) \citep{Curti+17}, and a similar calibration based on \RS\ $=\log( $[O \textsc{iii}]$\lambda5007/$H$\beta + $[S \textsc{ii}]$\lambda\lambda6717,6731$/H$\alpha $) \citep{Curti+20}. These calibrations were constructed by fitting polynomials up to 5th order between the line ratios and metallicities computed using the $T_e$ method for stacked spectra of galaxies observed by the Sloan Digital Sky Survey, and each relationship show scatters of the order of $\sigma=0.1$ dex. 

One limitation of the \ON\ and \RS\ diagnostics is that they have only been calibrated for a range of metallicities, from $\log (O/H) + 12 = 7.6$ to $8.9$. A negligible number of spaxels with a metallicity larger than this value are discarded (13 out of 2472 \Hii spaxels were discarded for the \ON\ diagnostic, and 36 out of 2518 for \RS). This cutoff can be seen in Figure \ref{fig:gradients}, especially for the \ON\ diagnostic. It is important to note this as a caveat; however, the small number of high-metallicity spaxels is unlikely to affect our results, and developing a correction term to extend the range of these diagnostics to larger metallicities in order to handle these few spaxels is beyond the scope of this study.
As an additional caveat, the \RS\ diagnostic is double-valued for metallicities below $\log (O/H) + 12 = 8$.  However, the spiral galaxies considered in this study are all large, local spirals, with metallicities far greater than this limiting value for all spaxels. Hence, this problem does not require further consideration.

For each of these diagnostics, we compute the error in the deduced metallicity of each spaxel from the uncertainty in each line flux $l_i$ using linear error propagation, and assuming the error in all line fluxes $l_i$ are independent:

\begin{equation}
    \sigma^2_Z = \sum \left(\frac{\partial Z}{\partial l_i} \sigma_{l_i}  \right)^2 
    \label{eq:Z_error}
\end{equation}
\subsection{DIG diagnostics}
\label{ssec:DIG-diagnostics}

\subsubsection{[S \textsc{ii}]/H$\alpha$}
\label{ssec:S2-diagnostic}
Our primary diagnostic for separating DIG-dominated regions from star-forming spaxels uses the [S \textsc{ii}]/H$\alpha$ line ratio, and the methodology of \citet{Kaplan+16}. \citet{Madsen+06} found that this line ratio is higher in DIG regions. Using this information, \citet{Blanc+09} created an equation that relates $C_{\text{H}\textsc{ii}}$, the fraction of H$\alpha$ emission originating from \Hii regions, from the [S \textsc{ii}]/H$\alpha$ line ratio of each spaxel:

\begin{equation}
    \frac{[\text{S} \,\textsc{ii}]}{\text{H}\alpha} =  C_{\text{H}\textsc{ii}}{\left( \frac{[\text{S}\, \textsc{ii}]}{\text{H}\alpha} \right)}_{\text{\Hii}} + \left( 1 - C_{\text{H}\textsc{ii}} \right) {\left( \frac{[\text{S}\, \textsc{ii}]}{\text{H}\alpha} \right)}_{\text{DIG}}
\end{equation}

To properly estimate $C_{\text{\Hii}}$ for each spaxel using rigorous statistical methods, three things would need to be known: (i) the distribution of the [S \textsc{ii}]/H$\alpha$ line ratio for DIG regions within each galaxy; (ii) the distribution of the [S \textsc{ii}]/H$\alpha$ line ratio for \Hii regions with no DIG contamination; and (iii) the prior expected distribution of $C_{\text{\Hii}}$ for each spaxel. In reality, none of these three distributions are known. Therefore, some assumptions must be made.

The method of \citet{Kaplan+16} makes two main assumptions: firstly, that the intrinsic distributions of [S \textsc{ii}]/H$\alpha$ are narrow, with little intrinsic scatter within each galaxy; and secondly, that the spaxels with the brightest H$\alpha$ flux in each galaxy have zero DIG contamination ($C_{\text{\Hii}}=1$), while the faintest spaxels are completely contaminated by DIG ($C_{\text{\Hii}}=0$). Under these two assumptions, ${\left( \frac{[\text{S} \textsc{ii}]}{\text{H}\alpha} \right)}_{\text{\Hii}}$ and ${\left( \frac{[\text{S} \textsc{ii}]}{\text{H}\alpha} \right)}_{\text{DIG}}$ are taken to be the median value of $[\text{S} \textsc{ii}] / \text{H}\alpha $ for the 100 spaxels with the brightest and faintest flux of H$\alpha$, respectively, for each galaxy. Following \citet{Poetrodjojo+19}, we take any spaxels with $C_{\text{\Hii}} > 0.9$ to be \Hii spaxels, and classify all other spaxels as DIG-dominated. For NGC 5236, this leads to $3664$ spaxels being classified as \Hii spaxels, and $19607$ being identified as having non-negligible DIG contributions.

Unlike $\Sigma_{\text{H}\alpha}$, [S \textsc{ii}]/H$\alpha$ is an intensive quantity, and does not suffer the conceptual limitations associated with line-of-sight projections detailed in \citet{Lacerda+18}. But it also agrees with the $\Sigma_{\text{H}\alpha}$ diagnostic, making it a reliable intrinsic proxy that is consistent with the $\Sigma_{\text{H}\alpha}$ method \citep{Blanc+09, Kaplan+16, Poetrodjojo+19}. A drawback of this method is that the decision to use 100 spaxels to classify the reference ratio values for DIG/\Hii regions is arbitrary. In the galaxies in our sample, this corresponds to $0.4-3.8\%$ of spaxels. We leave analysis on how this hyperparameter affects the realised DIG/\Hii distributions to dedicated studies of DIG classification.

\subsubsection{BPT diagnostics}
\label{ssec:BPT-diagnostics}

As an additional measure, we also classify spaxels as \Hii/DIG-dominated using the classical emission line ratio diagnostic diagrams, commonly known as BPT diagrams \citep{BPT, Veilleux+Osterbrock87}. This methodology is also used by \citet{Kumari+19} when constructing DIG-corrected versions of the \ON/\RS\ metallicity diagnostics. For consistency with this study, we adopt the same diagnostics to distinguish \Hii regions from DIG used by \citet{Kumari+19}, using the demarcation line published by \citet{Kauffmann+03} on the [O \textsc{iii}]/H$\beta$-[N \textsc{ii}]/H$\alpha$ diagram, and the \citet{Kewley+01} demarcation line on the [O \textsc{iii}]/H$\beta$-[S \textsc{ii}]/H$\alpha$ diagram. These diagnostics are some of the oldest, most well tested and trusted ways in which ionisation from \Hii regions is distinguished from other sources of ionisation and show quantitative agreement with other diagnostics \citep{Lacerda+18}. While they are traditionally used to classify galaxies and not regions within galaxies, these diagnostics are applicable to all spaxels in IFU datacubes with sufficient S/N (see, e.g. \citealt{Davies+14a, Davies+14b, Kewley+19, Sanchez20}).

We note that these different diagnostics predict a different number of \Hii spaxels and DIG regions for each galaxy. For NGC 5236, using the N2-based BPT diagnostic, $7232$ spaxels are identified as \Hii spaxels, and $2008$ are classified as being DIG-dominated; however, when the S2-based BPT diagnostic is used, $8819$ spaxels are identified as \Hii spaxels, and only $307$ are classified as DIG-dominated. For this reason, we caution the reader that the classification of each spaxel as being \Hii/DIG dominated depends on the DIG diagnostic used. To overcome this dependence, analogous Figures to those presented in this work are shown for all DIG diagnostics in the supplementary material (available online).

\label{lastpage}
\end{document}